\documentclass[journal, letterpaper]{IEEEtran}

\usepackage{epsfig,endnotes}
\usepackage{graphicx}
\usepackage{balance}  
\usepackage{subfigure}
\usepackage{times}    
\usepackage{url}      
\usepackage{float}
\usepackage{tabularx}
\usepackage{paralist}
\usepackage{xspace}
\usepackage[usenames,dvipsnames]{xcolor}
\usepackage{amsmath}
\usepackage{stackengine,inlinenum}
\usepackage{comment}
\newcommand{\fig}{Fig.\ }

\ifCLASSOPTIONcompsoc
  \usepackage[nocompress]{cite}
\else
  \usepackage{cite}
\fi

\newcommand{\gamazon}{\textit{G-MTurk}\xspace}
\newcommand{\glab}{\textit{G-Sample+}\xspace}
\newcommand{\wave}{\texttt{Wave}\xspace}
\newcommand{\cir}{\texttt{Circle}\xspace}
\newcommand{\cf}{\texttt{Grab}\xspace}
\newcommand{\abc}{\texttt{\lq\xspace{}abc\rq}\xspace}
\newcommand{\sig}{\texttt{Sig}\xspace}
\newcommand{\ud}{\texttt{User-defined}\xspace}
\newcommand{\zoom}{\texttt{Zoom}\xspace}
\newcommand{\swipe}{\texttt{Swipe}\xspace}
\newcolumntype{C}[1]{>{\centering}m{#1}}

\newcommand{\figref}[1]{Figure~\ref{#1}}
\newcommand\tabhead[1]{\small\textbf{#1}}

\newcommand{\etal}{\mbox{\emph{et al. }}}

\begin{document}
\title{Which One to Go: Security and Usability Evaluation of Mid-Air Gestures}

\author{\IEEEauthorblockN{Wenyuan Xu\IEEEauthorrefmark{1},
Xiaopeng Li\IEEEauthorrefmark{2}, Jing Tian\IEEEauthorrefmark{2}, Yujun Xiao\IEEEauthorrefmark{1}, Xianshan Qu\IEEEauthorrefmark{2}, Song Wang\IEEEauthorrefmark{2} and
Xiaoyu Ji\IEEEauthorrefmark{1}}\\
\IEEEauthorblockA{\IEEEauthorrefmark{1}Zhejiang University \quad
\IEEEauthorrefmark{2}University of South Carolina, Columbia \\
}}


\maketitle

\begin{abstract}
With the emerging of touch-less human-computer interaction techniques and gadgets, mid-air hand gestures have been widely used for authentication.  
Much literature examined either the usability or security of a handful of gestures. This paper aims at quantifying usability and security of gestures as well as understanding their relationship across multiple gestures. To study gesture-based authentication, we design an authentication method that combines Dynamic Time Warping (DTW) and Support Vector Machine (SVM), and conducted a user study with 42 participants over a period of 6 weeks. We objectively quantify the usability of a gesture by the number of corners and the frame length of all gesture samples, quantify the security using the equal error rate (EER), and the consistency by EER over a period of time. Meanwhile, we obtain subjective evaluation of usability and security by conducting a survey. By examining the responses, we found that the subjective evaluation confirms with the objective ones, and usability is in inverse relationship with security. We studied the consistency of gestures and found that most participants forgot gestures to some degree and reinforcing the memorization of gestures is necessary to improve the authentication performance. Finally, we performed a study with another 17 participants on shoulder surfing attacks, where attackers can observe the victims multiple times. 
The results show that shoulder surfing does not help to boost the attacks. 
\end{abstract}



\section{Introduction}
\label{sec:introduction}

The proliferation of various gesture capturing devices (e.g., touch screen and depth sensors) has enabled user-friendly ways to operate computers 
as well as to \textit{authenticate users}. Essentially, such gesture-based authentication is behavioral biometrics. Compared with traditional methods (e.g., passwords, tokens, or physiological biometrics),  gesture-based authentication has several advantages and is believed to be resistance to shoulder surfing, password thieves, or token loss. Not surprisingly, much work has been devoted into gesture-based authentication, and researchers have studied both contact-based and mid-air gestures. 
%
%
The contact-based gestures are harvested while users touch I/O devices physically. 
In comparison, mid-air gestures require no physical contact of devices, and thus 
can eliminate smudge attacks~\cite{Aviv:woot10}, avoid bacteria propagation, and allow scenarios where touch is impossible (e.g., in a clean room~\cite{cleanroom}). 
In light of these advantages, in this paper, we investigate the security and usability of mid-air gestures.

\begin{figure}[t]
\centering
\vspace{0mm}
\begin{tabular}{cc}
\small
{\includegraphics[bb=0 23 596 818, width=0.34\columnwidth]{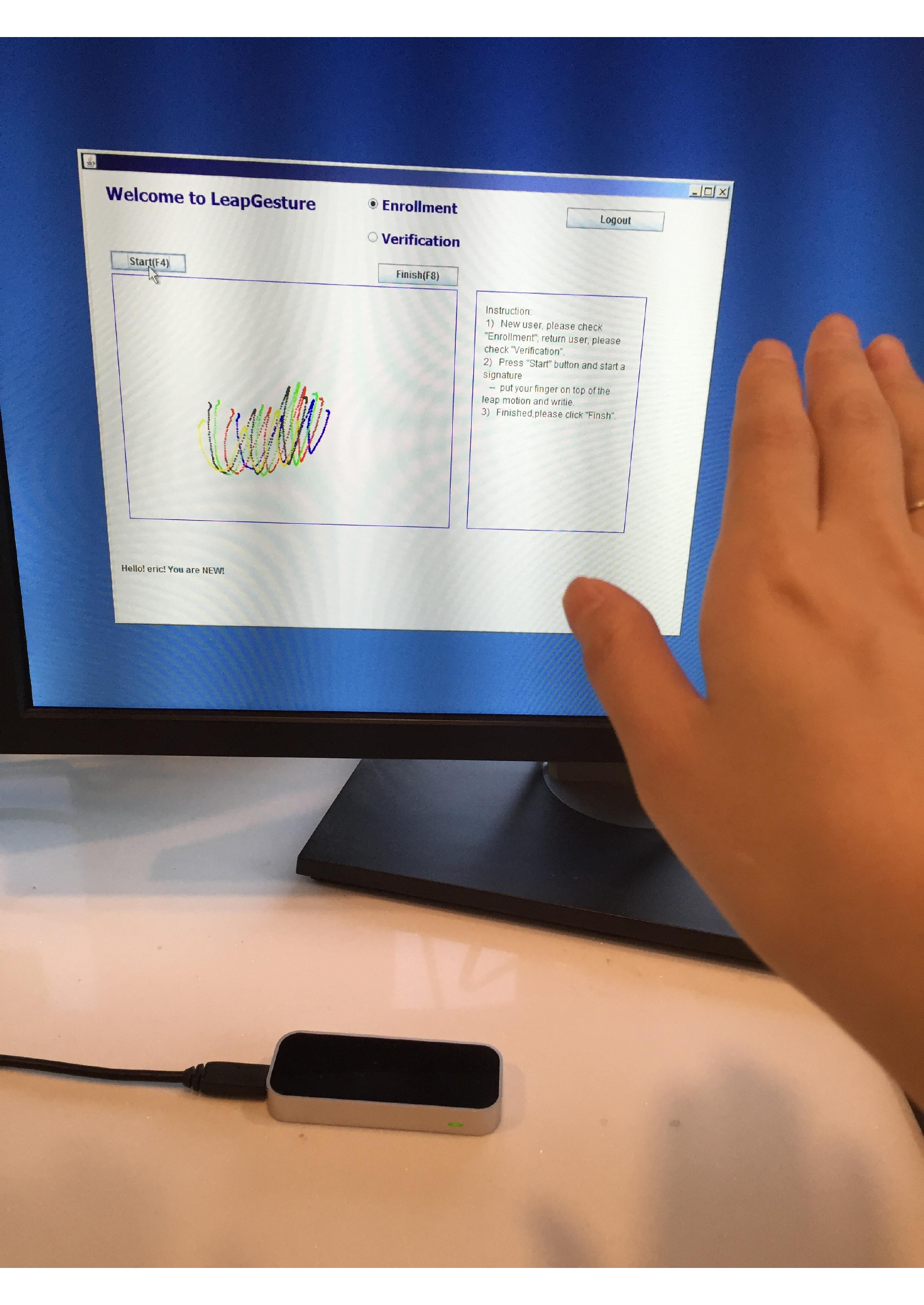}} 
& 
{\includegraphics[bb=0 23 596 818, width=0.34\columnwidth]{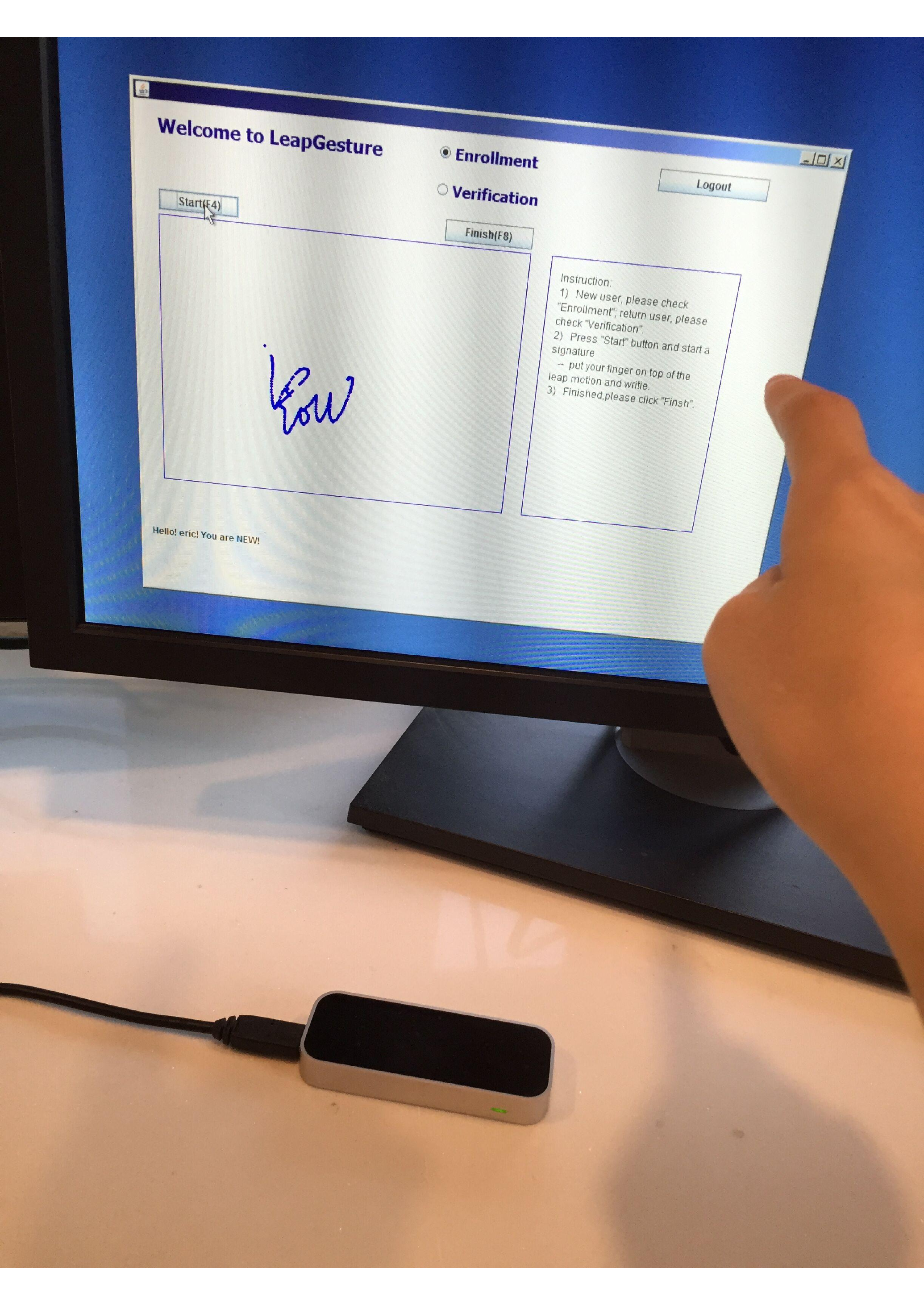}} 
\end{tabular}\vspace{0mm}
\caption{{Illustration of using various gestures as authentication inputs. Left picture shows waving with multiple fingers and right picture shows writing a \texttt{You} with one finger.}\vspace{-4mm}}
\label{fig:device}
\end{figure}


Already, researchers have proposed and studied various mid-air gestures for authentication, which, to name a few, include signature gestures captured by a Leap Motion controller~\cite{Nigam15:LeapSigVeri}, two `upward' hand movements~\cite{Midair_Authentication_Gestures_ICMP14},  simple gestures, such as drawing shapes, symbols, digits, etc., captured by a web camera with a short range depth sensor~\cite{airauth_chi14_abstract}. These works provide insights towards designing gestures for authentication. However, they either focused only on the security of mid-air gestures or performed preliminary user study on a limited number of gestures over a short period. 

It is so far unclear, does a complicated gesture always map to a higher level of security? Will a complicated gesture encompass larger variance and cause inconsistency in identifying a user? Does a complicated gesture represent poor usability, e.g., it takes long time to perform and is difficult to remember? Does gesture-based authentication share the same dilemma of passwords: what is secure is difficult to remember?  Given a gesture, can we provide quick feedback on its security level and thus assist in choosing a better gesture? This paper aims at answering these questions. 
In particular, we first selected a collection of representative mid-air gestures and user-defined gestures with the goal of exploring the trade-off between usability and security. We quantify security and usability of each gesture by using both \emph{objective} metrics that are calculated based on gesture samples and \emph{subjective} metrics derived from a survey. Using the gesture samples collected by $42$ users over 6 weeks and the survey responded by them, we managed to show that the quantitative metrics match with subjective perception of users and thus can be used to quantify the security and usability of gestures. Since we discovered that the usability and security are in inverse relationship, we can use the quantitative metrics of usability, i.e., the number of corners and the length of a gesture, to quantify the security of the gestures and provide quick feedback of a gesture. We summarize our contributions as below. 

\begin{itemize}
\item We proposed a set of metrics to quantify the usability and security of a gesture, which include objective metrics that are calculated based on gesture samples and subjective metrics derived from a survey.
\item We proposed an authentication method that combines a template-based method (DTW) and a machine-learning based classifier (SVM). The combined method can handle large spatial-temporal variations of a gesture by using a small number of training samples. 
\item We conducted two studies to quantify the gesture's security and usability: an objective evaluation by authenticating gesture samples collected over 6 weeks and a subjective evaluation by gathering well-designed questionnaires from users.
\item Our studies indicate that usability and security are in inverse relationship and thus we can utilize simple metrics (the number of corners and frames) to quantify the security of a gesture for quick feedback. In addition, our study suggests that repeated performing a gesture can improve users' perception of usability and help improve the consistency of gestures. 
\item Our study on shoulder surfing shows that hand gestures are hard to mimic and shoulder surfing attack is not a main thread to our authentication system.
\end{itemize}

\section{Overview}
\label{sec:overview}
In this section, we overview our problem definition and define metrics to quantify security, usability, as well as consistency. 

\vspace{-3mm}
\subsection{Problem}

Numerous gestures have been proposed to authenticate users, yet little has been done to compare their performance in terms of security and usability. 
This paper aims at filling in the blank by quantifying the security and usability of different mid-air gestures. We quantify security and usability of each gesture 
by using both objective metrics that are calculated based on gesture samples and subjective metrics derived from a survey. 
In particular, this paper tries to answer the following questions. 
\begin{itemize}
\item Security question: Given a set of gestures, which gesture maps to the best security level, i.e., it yields the best accuracy of authentication?
\item Usability question: Given a set of gestures, which one is the easiest to use and the most acceptable to users? How to quantify the usability purely using the statistics of gestures? 
\item Security vs. Usability: What is the relationship between security and usability when using mid-air gestures for authentication? 
Does gesture-based authentication share the same dilemma of passwords: secure gestures are more difficult to be remembered?
\end{itemize}

\vspace{-3mm}
\subsection{Security}

In the context of gesture-based authentication, we define security from the aspects of \textit{distinctness} and \textit{resilience to attacks} --- i.e., \textit{shoulder surfing attack}. 
A secure gesture should contain \emph{distinct} biometric information that suffices user authentication, i.e., even if two users perform the same gestures, their gesture samples should be distinguishable. In addition, a secure gesture should be resilient to attacks. Since much work claims that mid-air gestures are robust against should surfing attacks without validation, we focus on such attack.

\textbf{Metrics.} 
we use \textit{Equal Error Rate (EER)}, which is the value where the false rejection rate equals to the false acceptance rate, to quantify distinctness. In addition, we obtain users' subjective perception of security by conducting a user survey. Details are discussed in Section~\ref{sec:security}.

We use \textit{precision} and \textit{recall} to analyze the performance of each gesture password for defending against shoulder surfing attack. Precision is the percentage of honest users out of all the users that have passed verification, and it describes how cautious the system is to accept a user. Formally, $Precision = \frac{\sum_{i=1}^m tp_i}{\sum_{i=1}^m tp_i + \sum_{i=1}^m fp_i}$. Recall is the percentage of the honest users that have been granted access out of all honest users, and it affects the user experience.  Formally, $Recall = \frac{\sum_{i=1}^m tp_i}{\sum_{i=1}^m tp_i + \sum_{i=1}^m fn_i}$.


\vspace{-3mm}
\subsection{Usability}  

Motivated by the standard ISO 9241-11~\cite{ISO9241}, we define the usability of a gesture by considering  its efficiency, satisfaction, and learnability. Efficiency describes the resources required from users for successful authentication. Satisfaction reflects the comfort and acceptability of using the gesture, and learnability is defined by the ``time of learning'', i.e., how easy is it for users to pass the gesture-based authentication at their first attempt~\cite{Nielsen}? 

\textbf{Metrics.} To objectively quantify the efficiency and satisfaction of a gesture, we calculate \textit{the average length of the gesture samples} (i.e., how long does a user perform the gesture) and \textit{the average number of corners} in the gesture samples (i.e., the number of sharp turning points in the gesture). Intuitively, the longer it takes to perform a gesture or the more corners in a gesture, the less convenient the user feels 
and the poorer the usability. In addition, usability is subject to how users perceive. Thus, we also conducted a comprehensive user survey on the usability of each gesture. Details are discussed in Section~\ref{sec:usability}.

\vspace{-3mm}
\subsection{Consistency} 
Consistency (aka. memorability) can affect both security and usability, and thus we study consistency by itself.  An ideal gesture should be consistent over time with little memorization requirements:  when users return for authentication after a period of time since the last try, they can still provide gestures that contain the same biometric information as the ones that were initially enrolled for authenticating them. The more consistency, the better security performance and the less effort to pass authentication. 


\textbf{Metrics.} We quantified consistency from three aspects: \begin{inlinenum} \item the variances of each gesture over time (i.e., the frame number and corner number ) \item the EER of the gesture samples over a period of time (in our case, 6 weeks) \end{inlinenum}.
\section{System Design}
\label{sec:method}

We design an authentication system based on a Leap Motion controller, which is a 3D motion sensor and can track the motion of human hands as well as all ten fingers in the 3D space. We define a gesture sample as one measurement that contains a complete gesture, i.e., $N$ frames reported by Leap Motion. 
We  develop a program written in Java that integrated Leap Motion's SDK 2.0v~\cite{leap_Doc} for collecting gestures. After collecting the gesture data, we build a classifier which combines two algorithms (DTW and SVM) to distinguish users. 

In this section, we introduce the candidate gestures, feature selection and the classifier of our authentication systems. 

\vspace{-3mm}
\subsection{Gesture Selection}
\label{sec:gesture}

Many types of gestures have been studied in prior work, either in the context of gesture recognition or user authentication. These gestures include but not limited to swipe, zoom in/out, pan and scroll, point, and rotate either on touch screens or in the air~\cite{Freeman, SaeBaeCHI2012, soups14-paper-xu}; mid-air wave~\cite{CHI2014_wavetome}; mid-air signatures~\cite{TianQXW13:NDSS13}, etc. 
Covering every possible type of gestures is difficult, and thus 
we select a few popular gestures that are used for operating computing system and controlling home appliances (e.g., smart TVs) and/or have been studied specific for authentication, i.e.,~\swipe, \wave, \zoom and \cf and choose drawing gesture~\cir, writing gesture \abc and user-defined signatures (\sig) as they are studied for authentication purpose. Finally, we let each user define a gesture to reflect his/her preferences that are not included in the aforementioned gestures, we call it \ud gesture.

\begin{table}[t]
\centering
\scriptsize
\vspace{-0mm}
\caption{Illustration of the selected gestures. Each color corresponds to the trajectory of one finger. }
\label{tab:gesture_illustration}
 
 \begin{tabular}{ c  c | c  c  }
 \hline
    
     {\includegraphics[bb=0 21 298 189, width=0.16\columnwidth]{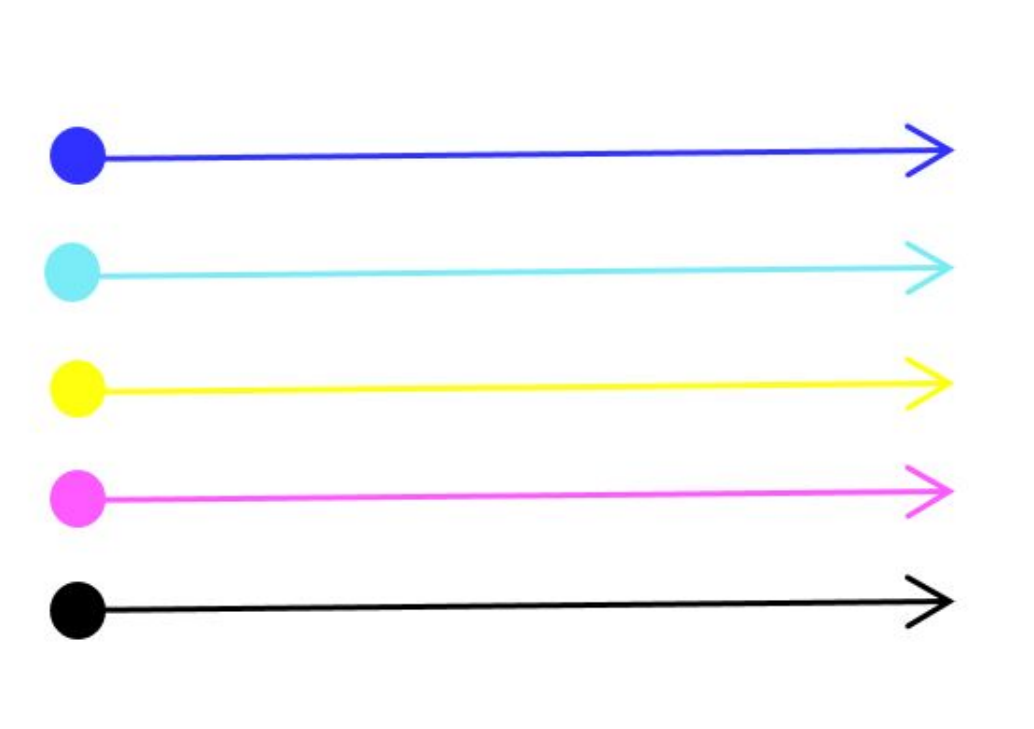}} 
    & {\includegraphics[bb=0 23 298 187, width=0.16\columnwidth]{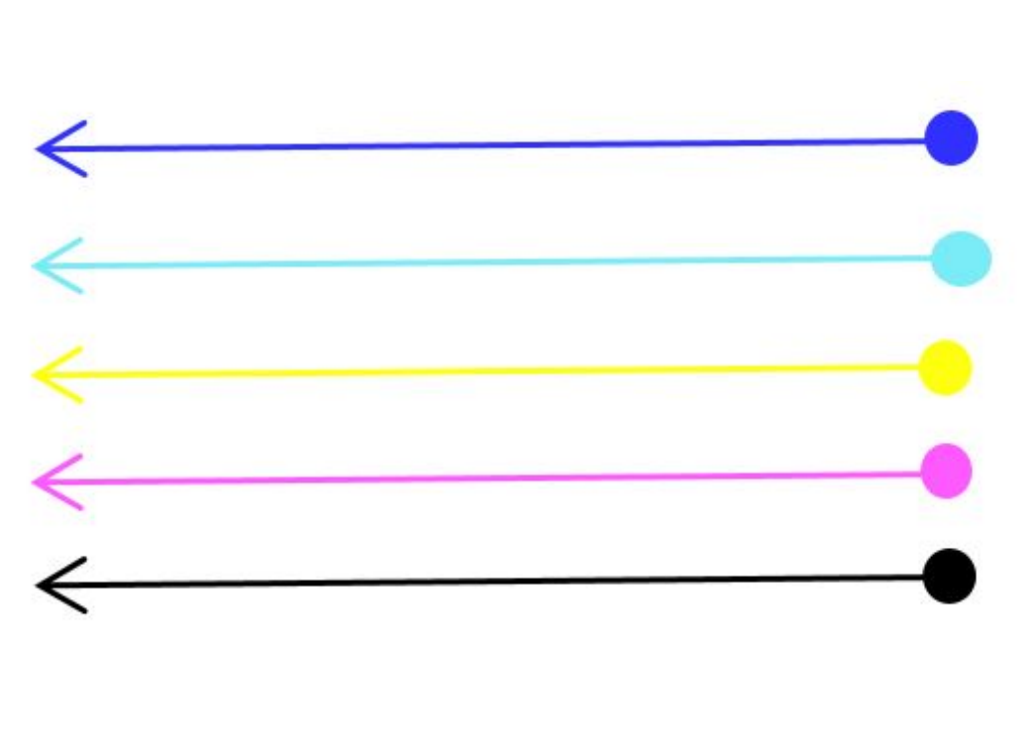}} 
    & {\includegraphics[bb=0 7 298 202, width=0.16\columnwidth]{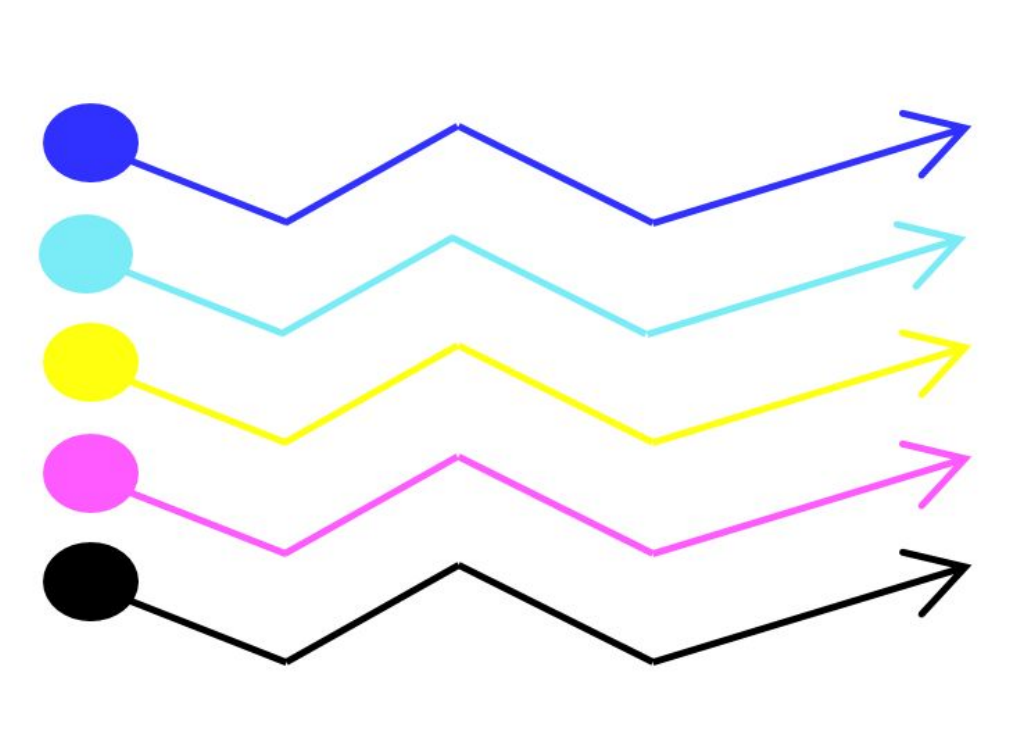}} 
    & {\includegraphics[bb=7 0 291 210, width=0.16\columnwidth]{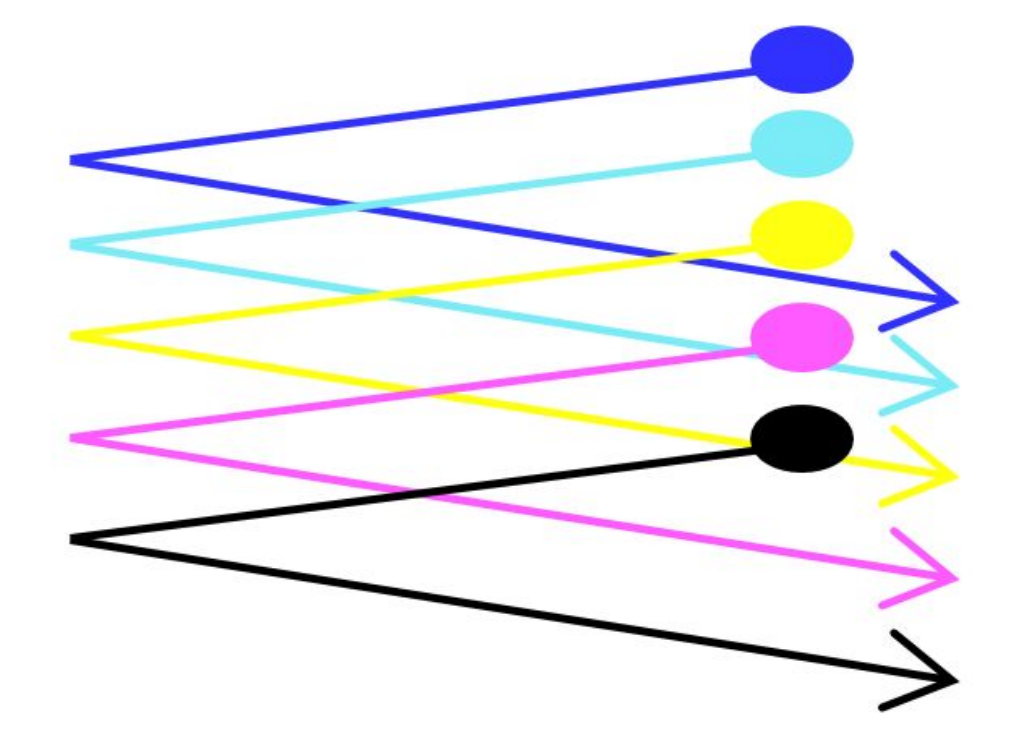}} \\
    \multicolumn{2}{c|}{\swipe}    &   \multicolumn{2}{c|}{\wave}  \\ \hline
      {\includegraphics[bb=18 0 279 210, width=0.16\columnwidth]{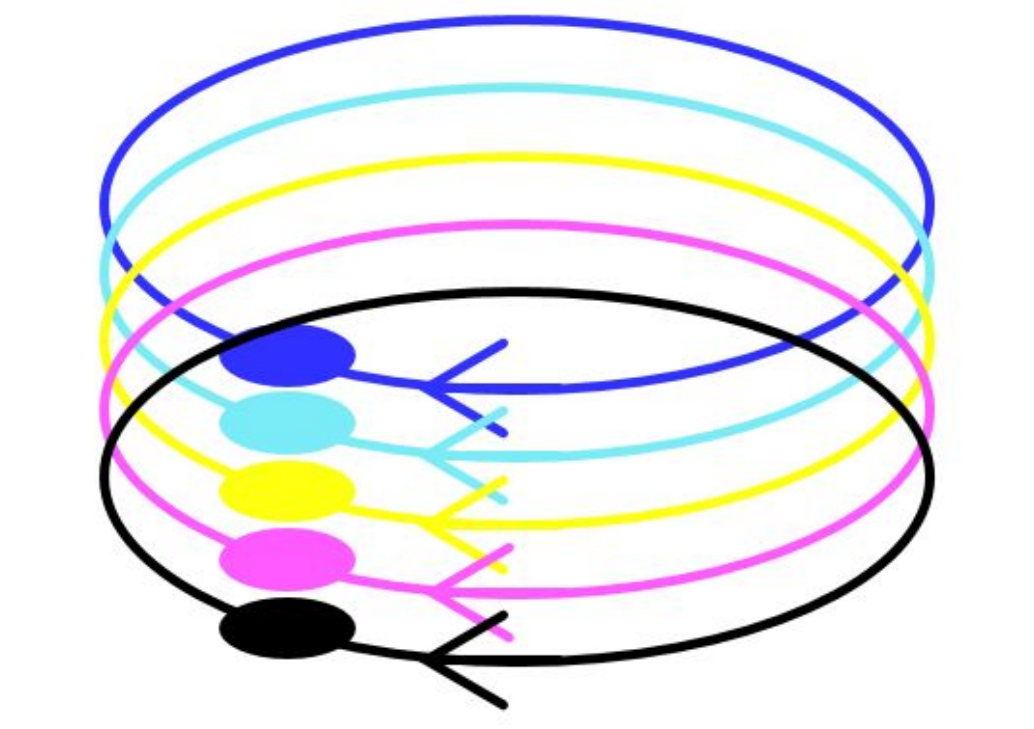}} 
     & {\includegraphics[bb=1 0 297 210, width=0.16\columnwidth]{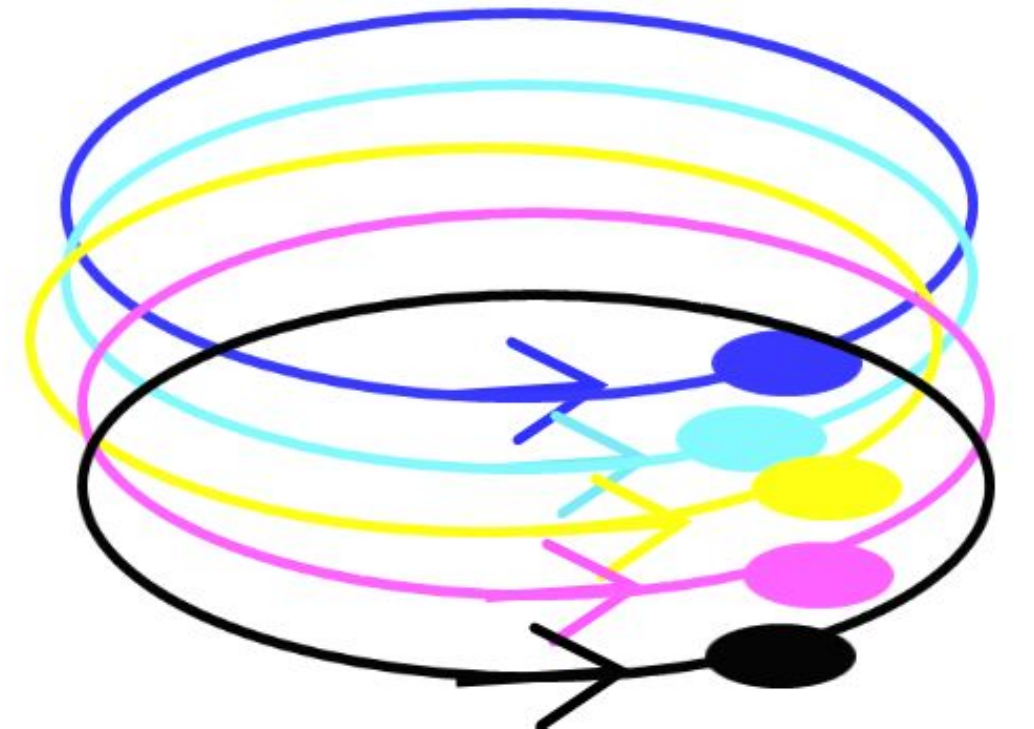}} 
       & {\includegraphics[bb=0 2 298 207, width=0.16\columnwidth]{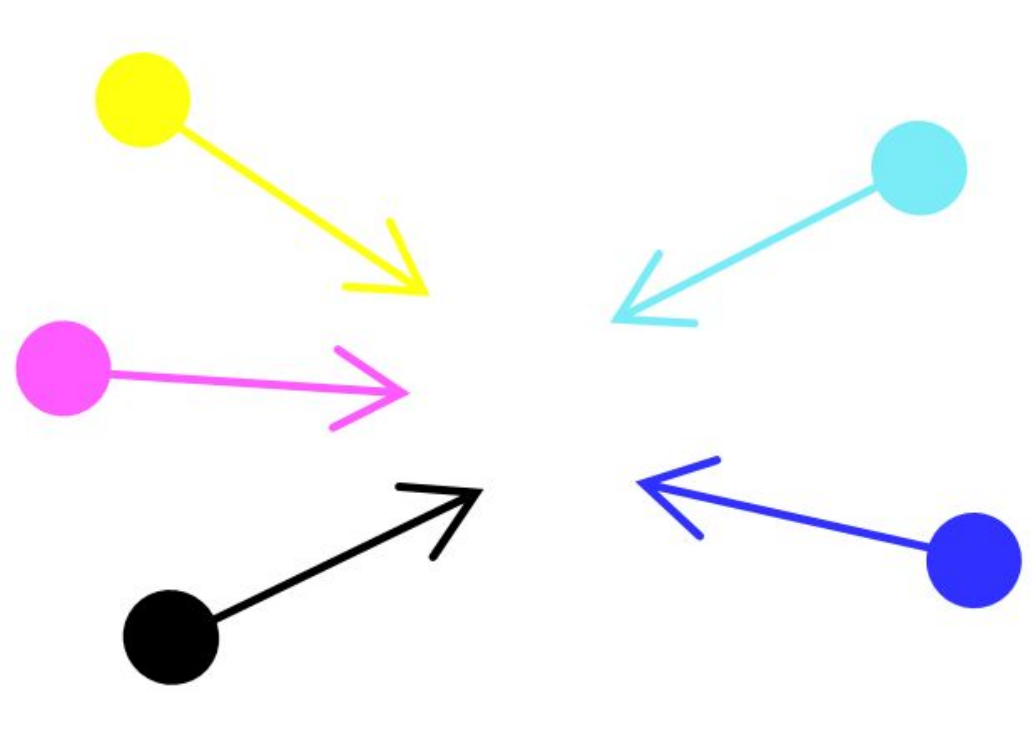}}  
       & {\includegraphics[bb=0 16 298 193, width=0.16\columnwidth]{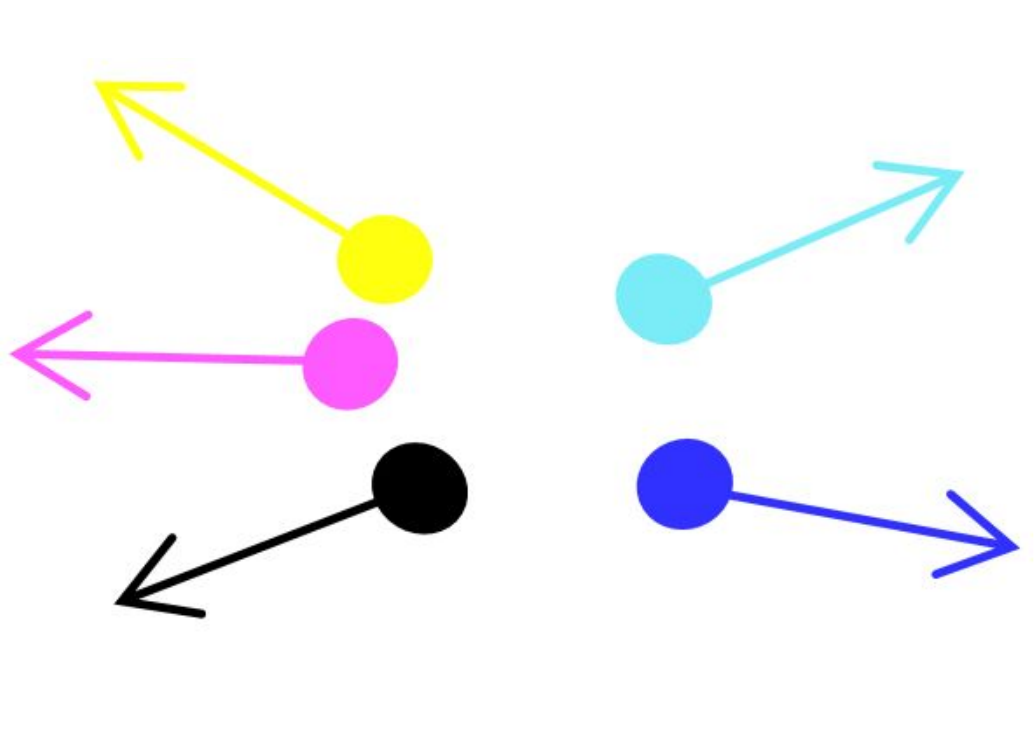}}  \\ 
 
   \multicolumn{2}{c|}{\cir}  &   \multicolumn{2}{c}{\zoom } \\ \hline
%
    {\includegraphics[bb=0 35 298 175, width=0.16\columnwidth]{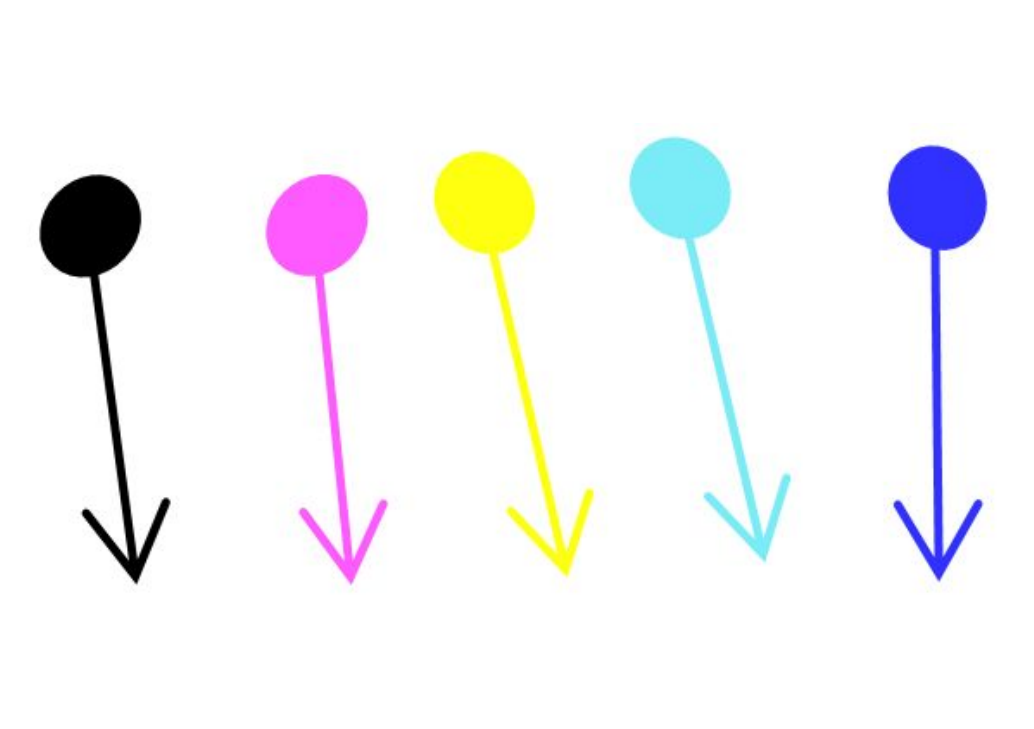}} 
    & 
    {\includegraphics[bb=0 15 298 195, width=0.16\columnwidth]{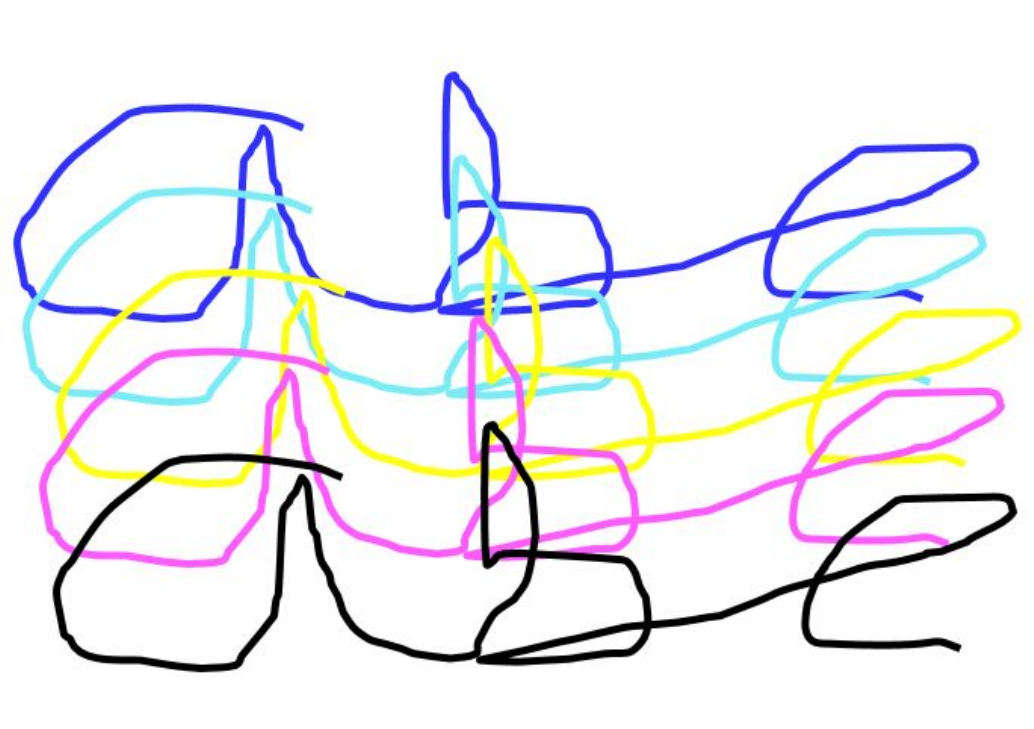}} 
     & {\includegraphics[bb=0 9 298 201, width=0.16\columnwidth]{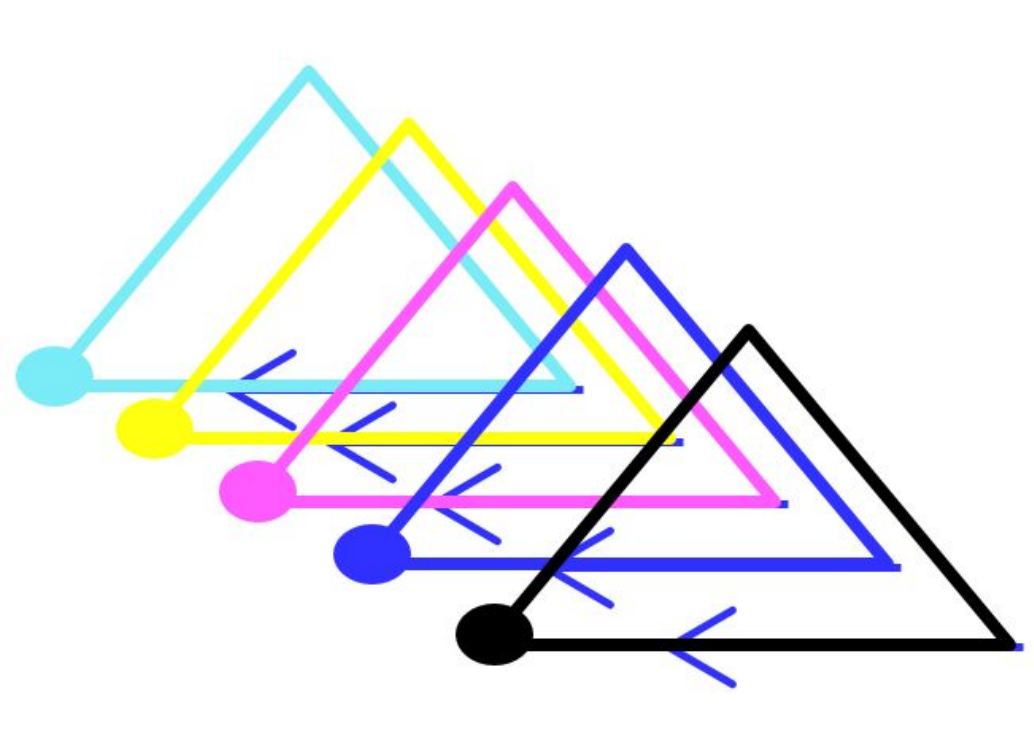}} 
       & {\includegraphics[bb=0 31 298 178, width=0.16\columnwidth]{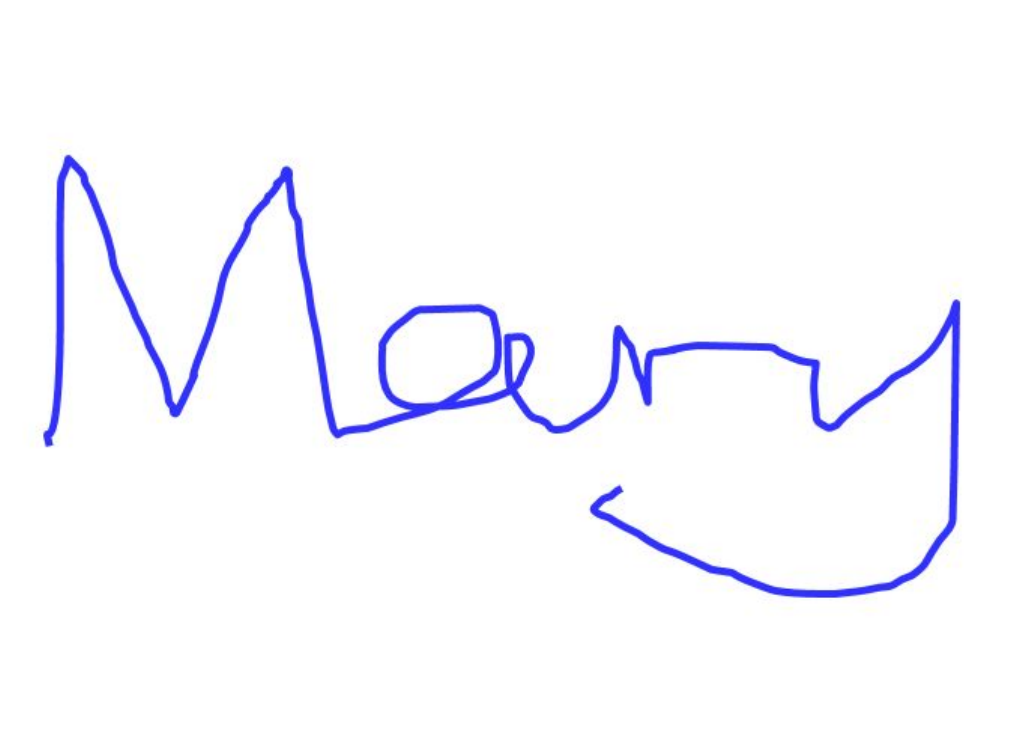}} \\
     \cf & \abc  &\ud  & \sig \\
 
  \hline
 \end{tabular}
 \vspace{-4mm}
\end{table}

In total, we select six pre-defined gestures 
and two user-defined gestures. 
We illustrate  all gestures in Table~\ref{tab:gesture_illustration}. 

\vspace{1mm}
\noindent \textbf{Pre-Defined Gestures}

1) \textbf{\swipe}. Users intentionally swipe his or her hand from one position to another position, and we define \swipe as a one way movement. Nowadays, swiping touch screens is a popular way to turn pages.

2) \textbf{\wave}. Users naturally wave their hand. We choose this gesture because we believe it is easy to perform and might lead to promising usability. Table~\ref{tab:gesture_illustration} illustrates two ways to wave: move up and down (i.e., mimicking ocean waves), or wave back and forth between left and right (i.e., waving hands to represent \texttt{Hello} or \texttt{goodbye}).

3) \textbf{\zoom}. Gestures \zoom in or out require to engage at least two fingers: \zoom either gathers finger-tips toward palm center or spreads out the finger-tips. Both gestures are commonly used for touch screen for changing the font size, showing/hiding a window, etc. We study \zoom in the context of 3D space.

4) \textbf{\cir}. Gesture \cir maps to the hand movement of drawing a circle when all five fingers are stretched out and towards the computer screen. The movement can be preformed clockwise or counter clockwise, and consists of one circle or multiple circles.

5) \textbf{\cf}. Gesture \cf is a quick, sudden clutch, starting with all fingers spread and ending at a fist.

6) \textbf{\abc}. Gesture \abc is the hand movement of writing a string `abc' with five fingers. It is chosen to test the gestures of writing letters/word. 


\vspace{1mm}
\noindent \textbf{User-Defined Gestures}

1) \textbf{\sig}. We let users sign their initials/first/full names that might be used for signing documents. Some of our participants used all five fingers to sign and some of them used only one finger.
  
2) \textbf{\ud}. We let each user to freely make one gesture that he/she believes to be secure and convenient for him/her. 

\vspace{-3mm}
\subsection{Authentication Algorithms}

Similar to most biometrics based authentication systems, we utilize supervised classifiers that require training to distinguish users. 
To effectively authenticate a user based on their mid-air gestures, we design an authentication algorithm that combines Dynamic Time Warping (DTW)~\cite{SaeBaeCHI2012} and Support Vector Machine (SVM)~\cite{light-svm} methods. 
As a template-based method, DTW is widely used to quantify similarity between samples and only requires a small number of templates. It allows nonrigid warping along the temporal axis and thus can tolerate differences in timing between gestures, i.e., a user may perform the same gestures at slower or faster speeds among trials. SVM is a popular machine learning algorithm that can handle more complex spatial-temporal variations of the same gestures at the cost of a large number of training samples. By combining the DTW and SVM methods, the proposed method can handle large spatial-temporal variations of a gesture by using a small number of training samples.

\textbf{SVM Training and Classification.}
We train a binary SVM classifier for each user with $T$ template samples. Specifically, given a system with $M$ users, to train the SVM classifier for user $i$, we take its $T$ template samples as the positive training samples and the template samples from other users as the negative training samples, i.e., $(M-1)T$ samples. For each training sample, we extract its $W$-dimensional DTW-based feature as the input to the SVM classifier, where $W=MT$.  
With the trained $M$ SVM classifiers, we can verify whether a new gesture sample $g_{test}$ is indeed performed by a user $u$ by \begin{inlinenum} \item computing the DTW distance between $g_{test}$ and all $W$ template samples to obtain the $W$-dimensional feature of the sample $g_{test}$, and \item input this $W$-dimensional DTW-based feature to the user $u$'s SVM classifier. \end{inlinenum} If the output of the SVM classifier is positive, the authentication of the user $u$ is succeeded. Otherwise, the authentication is failed.

%
%
%
%
%


To enroll a new user, we collect his/her $T$ template samples, add to the existing template samples, and retrain the SVM classifiers for all the users. In particular, the dimension of the sample feature will increase by $T$ when a new user is enrolled, i.e., $W=(M+1)T$.  Note that it is possible that users may  have various feature dimension of their SVM classifiers, depending on the sequence of their enrollment. In the verification stage, given a new gesture sample, we calculate its features, using which the trained SVM classifier verifies it.

\begin{table}[t]
\vspace{-0mm}
    \caption{Frame features from Leap Motion and DTW-based gesture-sample features.} 
    	
  \centering
  \small
    \begin{tabular}{l|l}
    \hline 
{Frame-based}  & \textbf{Hand}: \\
{feature} 
	& grab strength, and pinch strength,   \\
{for DTW}      & pitch, yaw, and roll, \\
		& palm width, and x, y, z axis of palm \\
		& x, y, z axis of arm and wrist.\\
       & \textbf{Finger}:\\
       & x, y and z positions of finger-tips,\\
       & x, y and z velocities of finger-tips, \\
       & x, y and z directions of finger-tips, \\
       & angles between consecutive frames.\\ \hline \hline
    {DTW-based} & \textit{DTW distance to $W$ template samples} \\
    {feature} &  sample $g'$s DTW distance to $W$ template samples \\
    {for SVM}  &   $\{g_{11}, g_{12}, \cdots, g_{1T}, g_{21}, g_{22}, \cdots, g_{2T}, \cdots, g_{M1},$ \\ 
    & $   g_{M2}, \cdots, g_{MT}\}$ \\ \hline   
    \end{tabular}
    \vspace{-4mm}
\label{tab:feature}
\end{table}

\vspace{-3mm}
\subsection{Feature Selection}
Based on our authentication algorithms, we extract two levels of features for DTW and SVM, respectively: frame features and DTW-based features. 

\textbf{Frame Features.} 
A raw data frame of Leap Motion contains $N$ frames with each frame containing 20 features for a hand and 11 features for each finger. Frame features consist of features directly from the raw data and the derived ones. The hand features include the following: grab strength and pinch strength, which describe the posture of the hand; pitch, yaw, and roll, which describe the angles of the hand around the $x-, y-$, and $z$-axes; palm width;  $(x, y, z)$ coordinates of palm, arm, and wrist, respectively; hand type, which indicates whether it is a left or right hand; 4	flags of gesture types, i.e., whether it is a circle, a swipe, a key tap, or a screen tap. 
The finger features include  the $(x, y, z)$ coordinates, the 3 dimensional velocity, and the moving directions of each finger tip, finger length, and finger width. 
Combing the features of the hand and its five fingers, we obtain 75 features on each frame from the raw data. In addition to these 75 features, on each frame we generate five new features based on finger features: the distance between finger tip positions in consecutive frames, two angles of finger-tip positions between consecutive frames in $x-y$ plane and $x-z$ planes, one angle in the 3D plane, and one curvature in the 3D plane~\cite{leap_Doc}. This way, we have 25 new features over five fingers and in total we obtain 100 features on each frame. 

\textbf{DTW-Based Features.} In the enrollment stage, we collect $T$ gesture samples for each of the $M$ enrolled anchor users. This way, we in total have $W=T\times M$ template samples $g_{ij}$, where $i$ indicates user $i$ 
and $j$ indicates the $ith$ template samples from each user. Given a gesture sample $g$, we extract a $W$-dimensional sample feature vector by computing and concatenating its DTW distances against all the $W$
template samples, by following a fixed order of $\{g_{11}, g_{12}, \cdots, g_{1T}, g_{21}, g_{22}, \cdots, g_{2T}, \cdots, g_{M1}, g_{M2}, \cdots, g_{MT}\}$. This $W$-dimensional sample feature 
is then used as the input to train and test the SVM classifiers. 

\textbf{Feature Reduction.} Since the $100$ features of each frame do not contribute equally towards verification, we select a subset of them to compute the DTW distance with the goal of maximizing the verification performance. To evaluate each feature, we use each of these 100 frame features to compute the DTW-based feature as mentioned above for training the SVM classifier and evaluating the average EER over all the users (to be discussed later).  We discard the frame features that produce an EER less than 50\%. 
Eventually, we kept $75$ frame features. To further boost the verification performance, we calculate the weight for each frame feature and use feature weights for computing the DTW distance between gesture samples.

\section{Data Acquisition}
\label{data}
To quantify the gesture's security and usability, we recruited 42 volunteers (32 males and 10 females) in two universities. Among the 42 participants, 40 people are between $18$ and $34$ years, majority of whom are college students, and 2 participants are between $35$ and $54$ years. They were asked to complete gesture collection over 6 weeks and finish a survey in the end. Among the 42 participants, 32 participants perform all types of gestures. To mimic the real scenarios where a user may only need to remember a few \ud (UD) gestures, the other $10$ users in this group only contribute to \ud gestures. When participants perform gestures, we encouraged them to perform in the most comfortable ways. Each participant was compensated a $\$20$  gift card after completion the whole experiment. 

\begin{table}[t]
\scriptsize
\vspace{-0mm}
\caption{Basic information on data collection.}
  \centering
\begin{tabular}{c|ccc}
    \hline
    \tabhead{Data} & \tabhead{$\#$Participants} & \tabhead{$\#$Samples} & \tabhead{Ave. days after} \\
    \tabhead{Batch No.} & \tabhead{excpet \it{UD only}} & \tabhead{ } & \tabhead{1st collection} \\ \hline
   
    \tabhead{1st } & 32 & 3400 & 0 \\
    \tabhead{2nd} & 32 & 2724 & 2 \\
    \tabhead{3rd} & 32 & 2567 & 5 \\ 
     \tabhead{4th} & 32 & 2519 & 8 \\
    \tabhead{5th}  & 31 & 2352 & 10 \\
     \tabhead{6th}  & 29 & 1904 & 12 \\ 
     \tabhead{7th}  & 28 & 1732 & 15 \\
     \tabhead{8th}  & 28 & 1693 & 17\\ 
     \tabhead{9th}  & 27 & 1490 & 24\\ %
     \tabhead{10th}  & 24 & 1297 & 27\\ 
     \tabhead{11th}  & 14 & 954 & 32\\
     \tabhead{12th}  & 13 & 885 & 37\\
     \tabhead{13th}  & 13 & 874 & 43\\ 
     \hline
 
  \end{tabular}   \vspace{-4mm}
  \label{tab:data}
\end{table}

Table \ref{tab:data} summarizes the information for each round of data collection. Each batch denotes that we collect the participants' gesture data for one time. In the first two weeks, the participants came to our lab three times per week and in the third and forth weeks, twice per week. For the last two weeks, the participants came to our lab three times in total. The time elapsed between two consecutive  data collections are more than 24 hours. In total, we collected 13 batches of data around 6 weeks.

\textbf{Pre-defined Gestures.} Most of the pre-defined gestures are commonly used on touch screens or pad, and users usually know how to perform them. Nevertheless, users have their own preferences of performing gestures.  For instance, a user may wave from left to right, and another user may wave from top to down. 


 

\textbf{User-defined Gestures.} 
1) {Gesture \ud}. Participants were encouraged to select one gesture that is secure for authentication and convenient to use. Among the 42 participants, 25 chose letter(s) and number(s). 17 participants chose to draw simple shapes, such as mathematical symbols, stars, and the combination of the above shapes or some other strange shapes. 

2) {Gesture \sig}. 
For convenience concerns, most participants did not sign their whole name, but just their initials, first names, or family names. Among all of them, initials are the most popular choices, which account for $70\%$. The average length (i.e., the number of English letters) of the collected signatures is $4.4$ with a maximum of 8 and minimum of 1. 


\textbf{Survey.} After finishing the gesture collection over six weeks, the participants were asked to answer a survey. The question answers use a 5-point Likert scale, i.e., 5 choices ranging from “Strongly agree” to “Strongly disagree”.  
The survey consists of three parts: (1) Part one asks for participants' background information, e.g., gender and age and their preferences on authentication mechanisms. 
(2) The second part of the survey consists of a set of $12$ questions (in Table~\ref{tab:survey}) for each gesture. 
We modified $10$ standard System Usability Scale (SUS)~\cite{Brooke96sus} questions to measure the usability 
of each gesture, and added $2$ more questions to measure the memorability. 
(3) In the end, we asked participants to rate the security level of each gesture if they are going to use these gestures as passwords.

\begin{table}
  \centering
  \scriptsize
  \vspace{-0mm}
  \caption{Questions for each gesture in the survey. \vspace{0mm}}
  \begin{tabular}{l|l}
    \hline
      &\tabhead{Question}  \\ \hline
1 & I would like to use this gesture frequently. \\ 
2 & I found it unnecessarily complex. \\
3 & I thought it was easy to use.\\ 
4 & I would need training to be able to use it. \\ 
5 & I found it would be performed smoothly.\\ 
6 & I think I cannot perform the same gesture every time. \\
7 & I would imagine that most people would learn to use it very quickly.\\
8 & I found it very cumbersome to use. \\
9 & I felt very confident using it. \\
10 & I needed to learn a lot of things before I could get going with this \\
& gesture.\\
11 & I can easily remember how to perform this gesture. \\
12& It is hard for me to recall this gesture after one week. \\
    	   \hline 
  \end{tabular}\vspace{-4mm}
  \label{tab:survey}
\end{table}

\section{Evaluation}
\label{sec:evaluation}

We objectively evaluated the security and usability of all the gestures by analyzing the collected samples, e.g., computing the number of corners, the number of frames, and the EER of each gesture. 
We also subjectively evaluated the security and usability of all the gestures by conducting a survey from all the users. In the end, we try to explore the relationship between usability and security from both objective and subjective perspectives.


\vspace{-3mm}
\subsection{Security Evaluation}
\label{sec:security}

We discuss distinctness in this section and  consistency in Section~\ref{sec:consistency} and shoulder surfing attack in Section~\ref{shouldersurfing}. In this section, we first summarize the subjective security results reported by participants and then quantify security objectively by EER. 

\subsubsection{Results from Survey Responses}

\begin{figure}[t]
\centering
\includegraphics[bb=85 17 680 354, width=0.74\columnwidth]{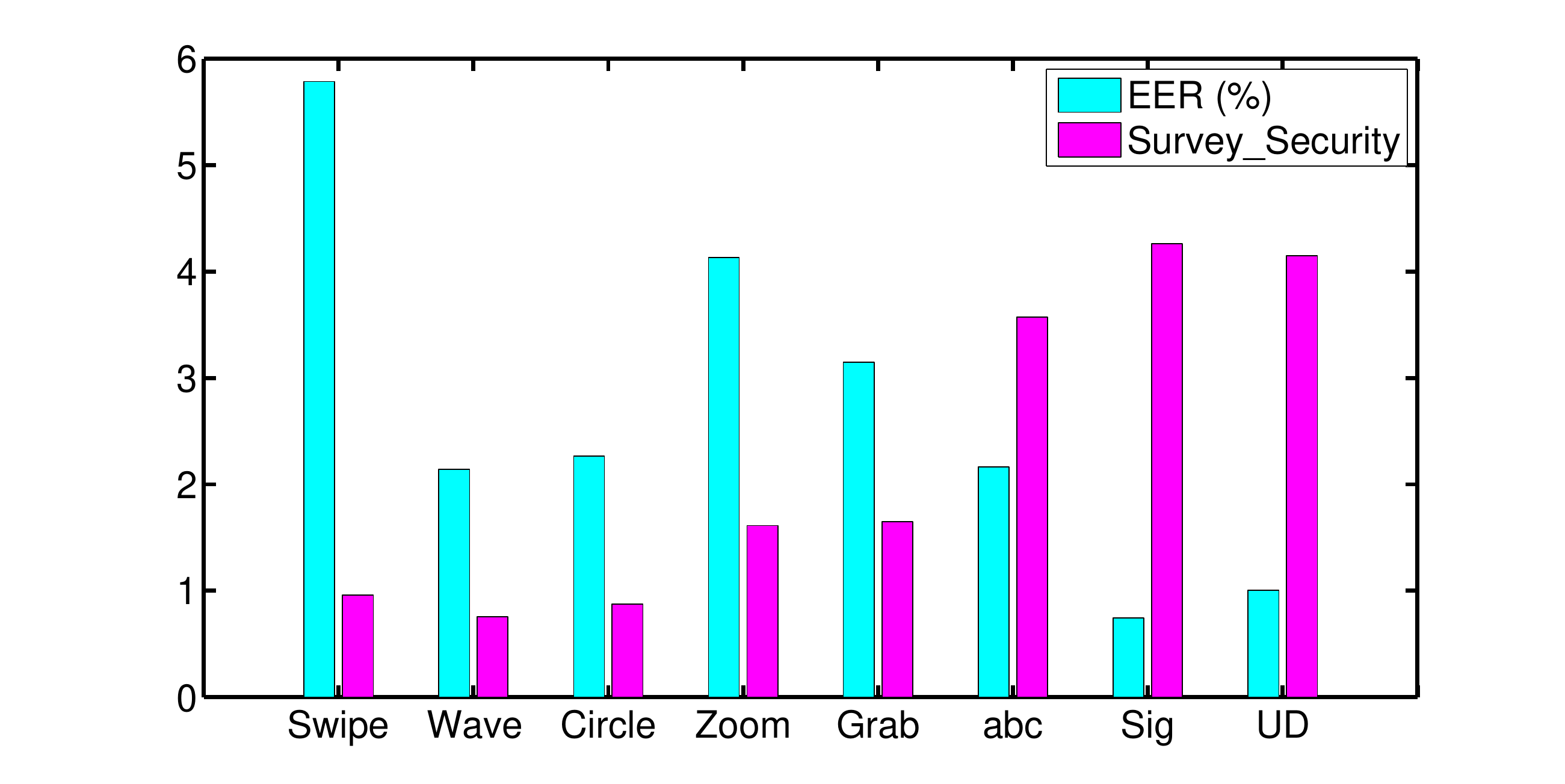}
\vspace{-2mm}
\caption{The security performance evaluated by survey responses from all participants and the EER from the authentication system.}\vspace{-4mm}
\label{fig:score_eer}
\end{figure}

The third part of the survey evaluates each gesture's security level if it is used as a password, and the question uses a 5-point Likert scale ``Least secure'' - ``Most secure''. In \fig \ref{fig:score_eer}, the pink bars show the score of security estimation derived from the survey responses. We first count the percentage of participants who chose ``Second secure'' and ``Most secure''. Then we divided the percentage values ($<100$) by 20 to fit the scale of EER ($<6\%$), which are shown in sky blue bars.


From the survey results in \fig \ref{fig:score_eer}, we have the following observations: 
\begin{inlinenum}
\item Pre-defined simple gestures \wave, \swipe, \cf, \zoom, and \cir are considered insecure.
\item Gesture \abc is considered to have the medium security level.
\item Gestures \sig and \ud are considered the most secure authentication gestures. 
\end{inlinenum}

\subsubsection{Results from Equal Error Rate}

To verify the overall performance of the system, we tested all the data with five folds of experiments. For each gesture type and each user, we randomly divided the data into a training set and a test set, and the overview performance is the average results of the multiple rounds experiments. In this experiment, we utilize the gestures collected in the first round, and we have $N$ users, and set the number of training samples for each gesture as $T = 4$. Therefore, the feature $f(g')$ for gesture $g'$ has $N\times4$ dimensions. To prepare a DTW-based feature for SVM classifiers, we need around $15$ms to compute features of a single testing gesture using a laptop with an Intel i7-2.8 GHz CPU and 8-GB memory.

The sky blue bars in~\fig \ref{fig:score_eer} show the average EER for each gesture. A smaller EER maps to a higher verification performance. The results show that: 

\begin{itemize}
\item The user-defined gesture \sig has the lowest error rate ($0.77\%$). Even participants may choose the same signature, the accuracy is still high as participants should have different writing styles.
\item  Although we observed that most participants chose simple movements, the user-defined gestures \ud (UD) have error rate $1.01\%$, 
\item The gesture \abc has an error rate of $1.81\%$. Although all the participants wrote the same content, we can identify the owner of each sample. Compared with other pre-defined gestures, \abc is more complex and contains distinct biometric information of participants.
\item Other pre-defined gestures have error rates ranging between $2\%$ and $6\%$. 
\item  The subjective security evaluation matches with the objective EER. The smaller the EER is, the more secure the authentication is.
\end{itemize}

\begin{figure}[!tbp]
\centering
\scriptsize
\includegraphics[bb=12 9 420 206, width=0.80\columnwidth]{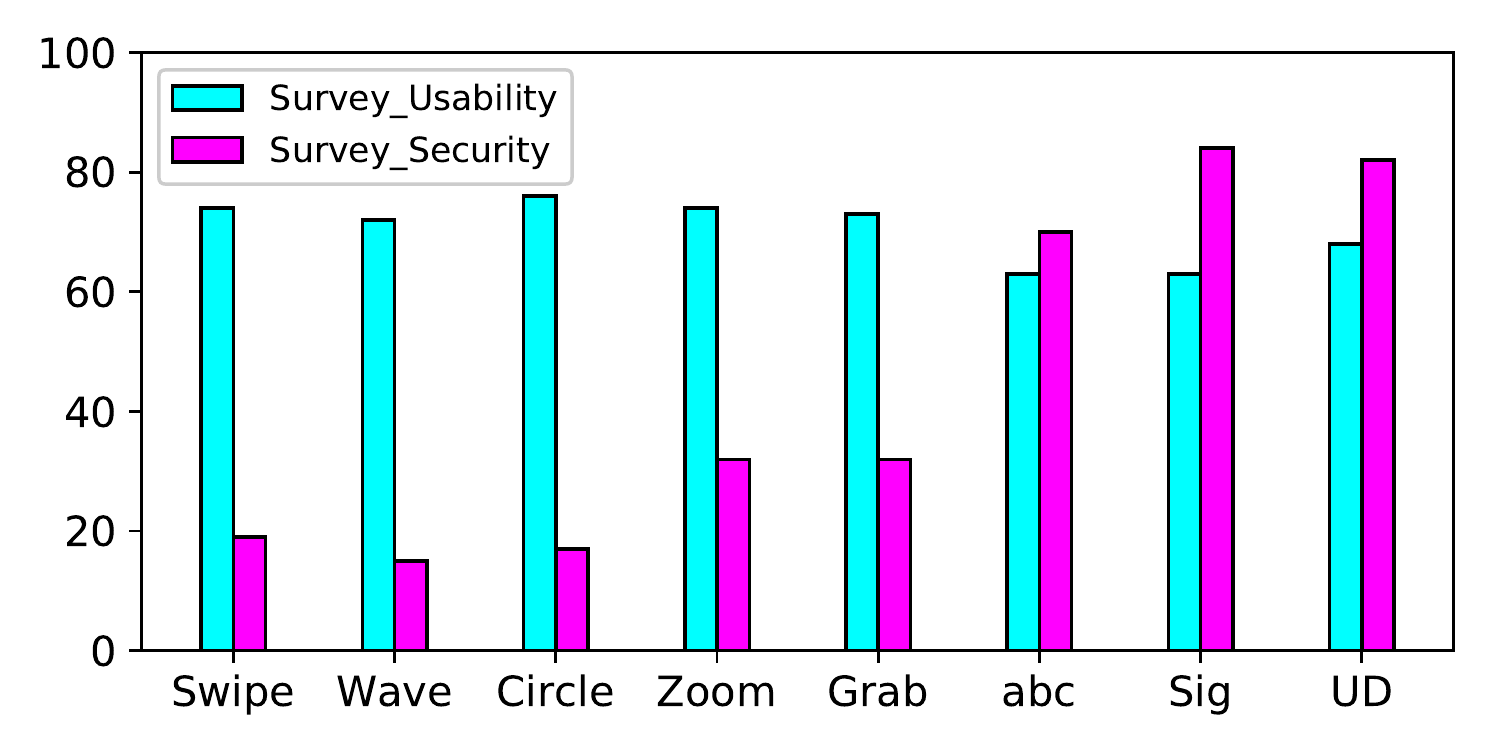} 
\vspace{-2mm}
\caption{{The results of participants' evaluation on both security and usability on each gesture. \lq\xspace{}100\rq  - best usability/security. \lq\xspace{}0\rq  - worst usability/security. }\vspace{-4mm}}
\label{fig:usa_sec_survey}
\end{figure}


\vspace{-3mm}
\subsection{Evaluation on Usability}
\label{sec:usability}
In this section, we first evaluate the usability of the authentication system from the survey responses. Then we evaluate the usability of each gestures from both the subjective aspect reported by the participants and the objective aspect quantified by the two metrics. 

The collected background information from the survey shows that the majority (90.5\%) of participants would like to use mid-air gesture for authentication if available, even including the older participants. 
The participants' responses further show that the reason of the acceptance include convenience, ease-to-remember, and security. Only few of them have concerns on security, worrying that mid-air gestures are not as accurate as typing a text password.

\subsubsection{Results from Survey Responses}


To evaluate the usability of each designed gesture, we calculate SUS scores using the responses of the first 10 questions for each gesture. Note that the SUS scores are references to compare participants' opinions among different gestures, i.e., the gesture with a higher score indicates a higher usability than the gesture with a lower score. The blue sky colors in \fig \ref{fig:usa_sec_survey} show the average score of participants for each gesture. We find that the SUS scores for simple predefined gestures \swipe, \wave, \cir, \zoom and \cf are close to or greater than 68  (above average), indicating participants are more willing to use gestures than unwilling\cite{68_SUS}. To analyze if different gestures has significant impact on perceived usability, a Repeated Measures Analysis of Variance (ANOVA) test was conducted using different gestures as independent variables and usability as the dependent variable. According to the result, there was not a significant effect of the gesture type on the perceived usability at the $p < 0.05$ level for the seven different gestures $[F(6, 175) = 0.69,\, p = 0.6613]$, which suggests that the participants consider all gestures have the similar usability.

\begin{figure}[t]
\scriptsize
\centering
\vspace{-0mm}
\begin{tabular}{ccc}
\subfigure[\cir]
{\includegraphics[bb=0 0 560 418, width=.25\columnwidth]{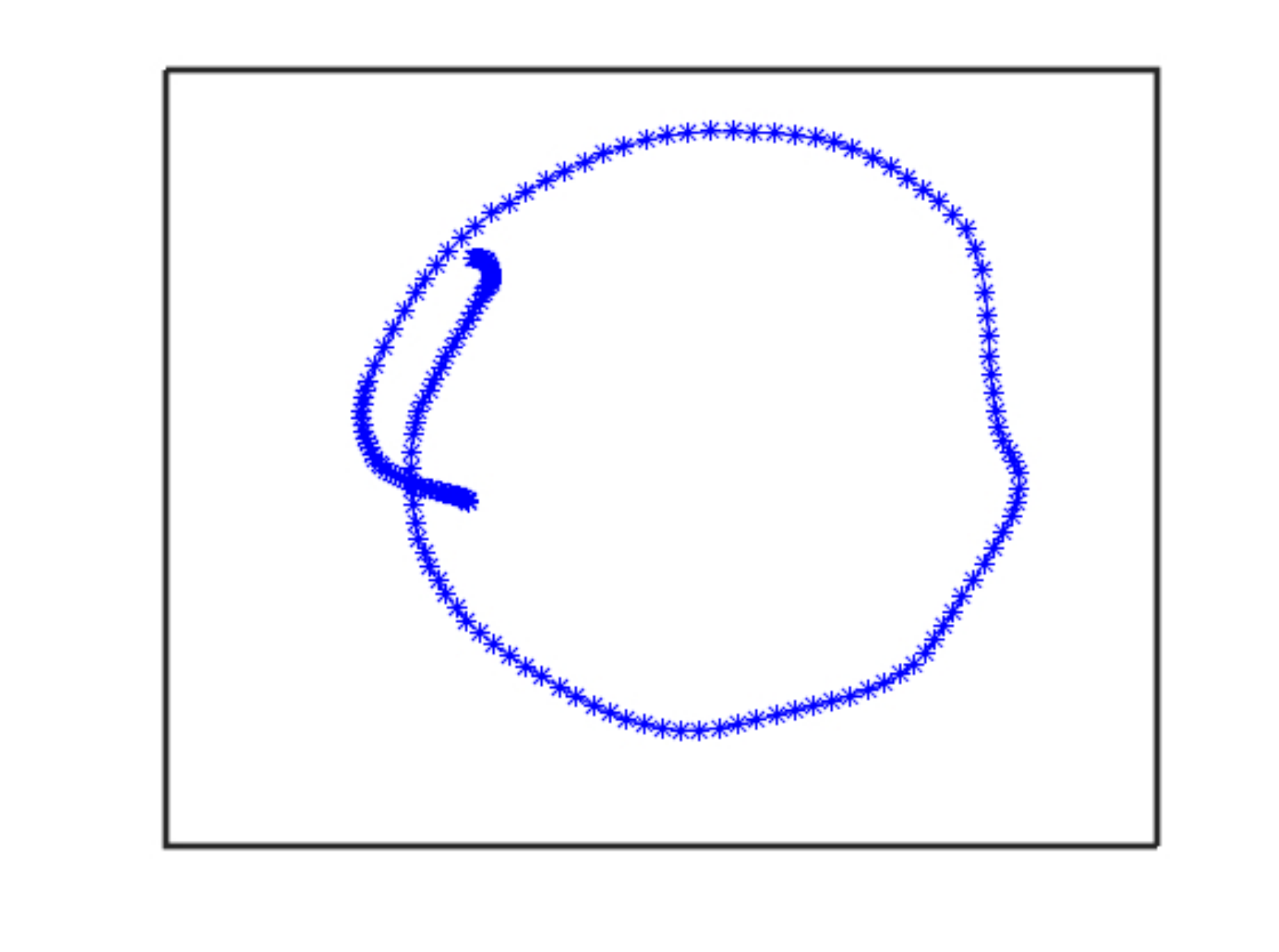}}
& \subfigure[\zoom]
{\includegraphics[bb=0 0 560 420, width=.25\columnwidth]{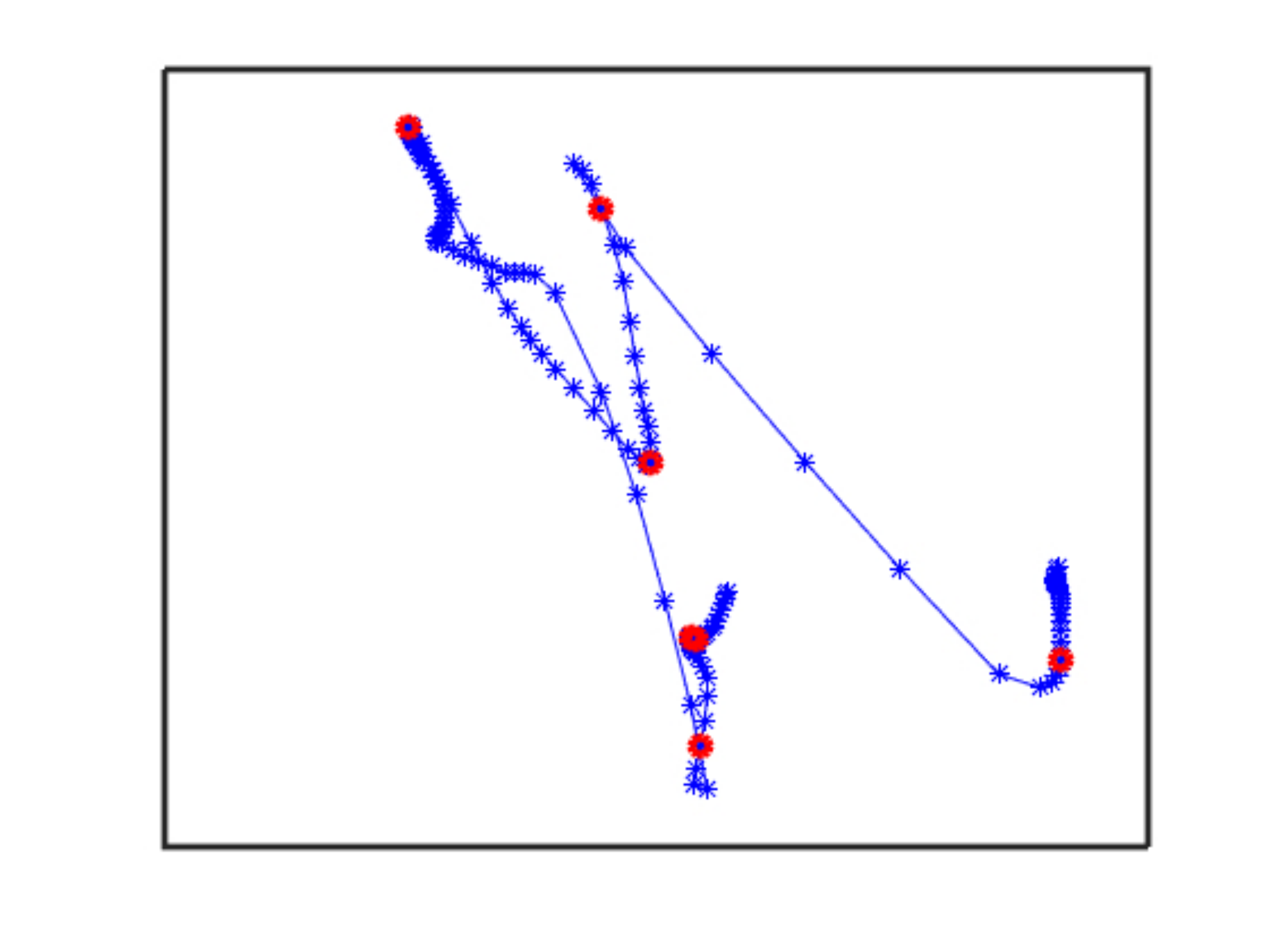}} 
&\subfigure[\cf]
{\includegraphics[bb=0 0 560 420, width=.25\columnwidth]{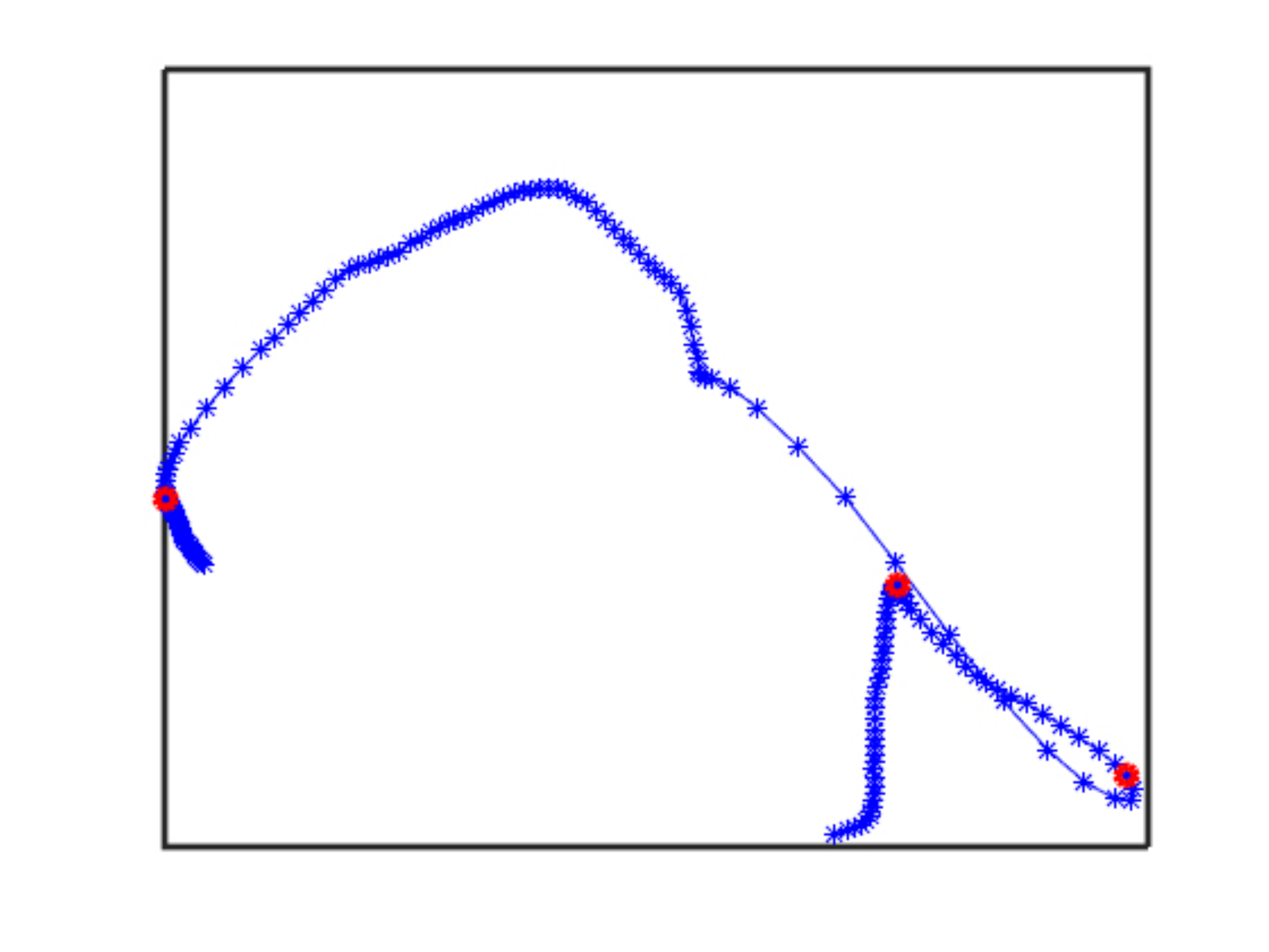}} \\
 \subfigure[\abc]
{\includegraphics[bb=0 0 562 420, width=.25\columnwidth]{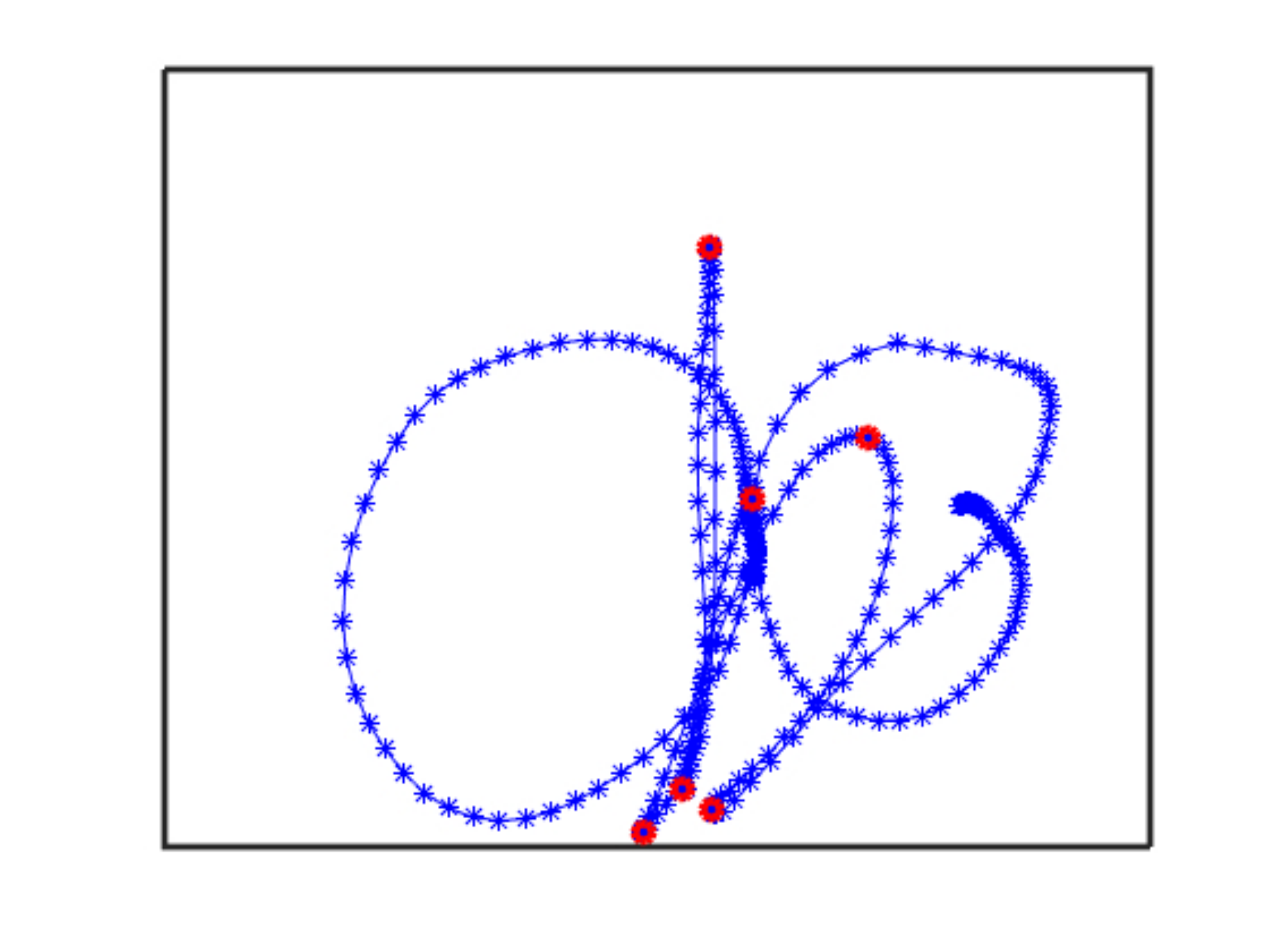}} 
& \subfigure[\texttt{UD}]
{\includegraphics[bb=0 0 560 436, width=.25\columnwidth]{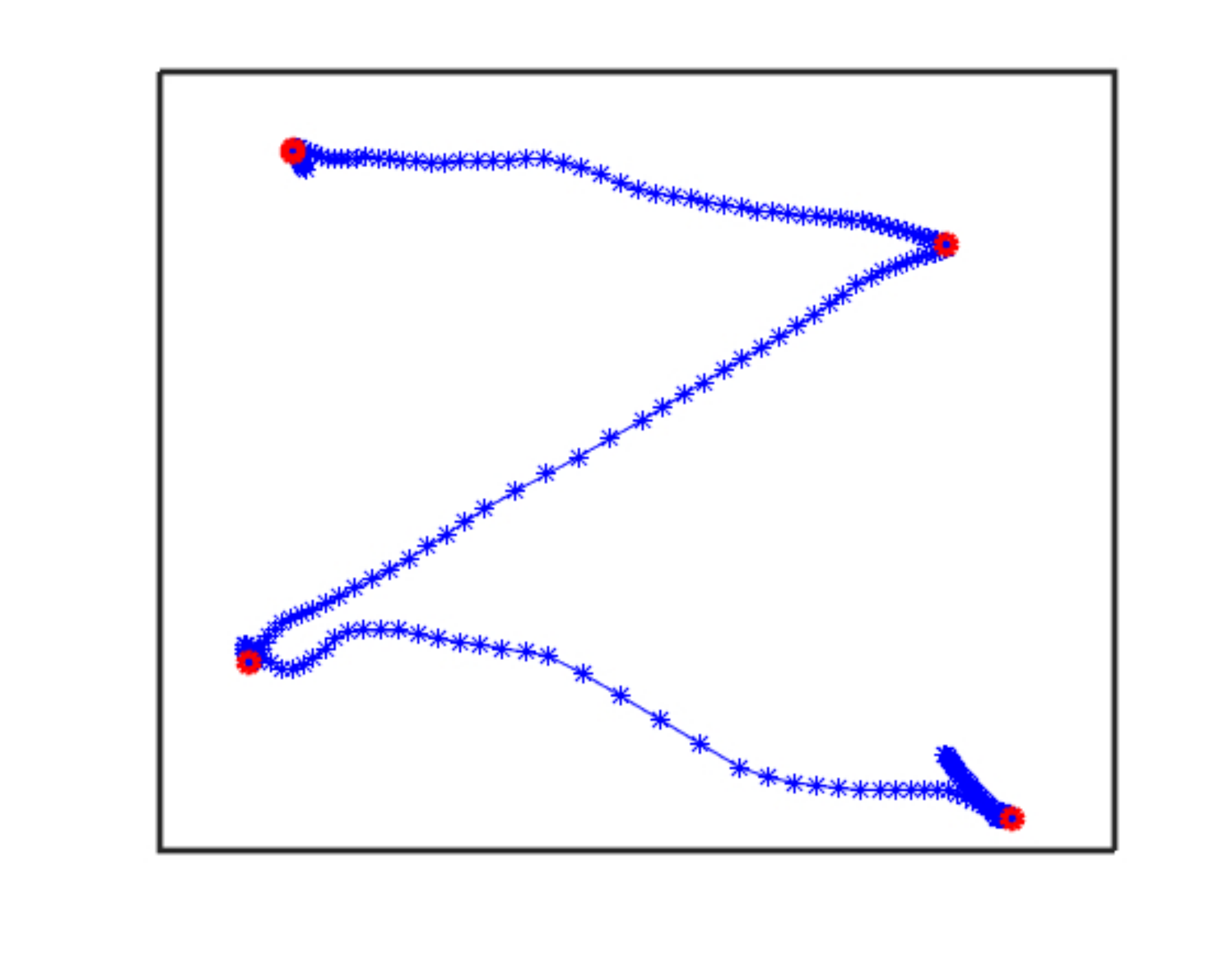}}
& \subfigure[\sig]
{\includegraphics[bb=0 0 560 420, width=.25\columnwidth]{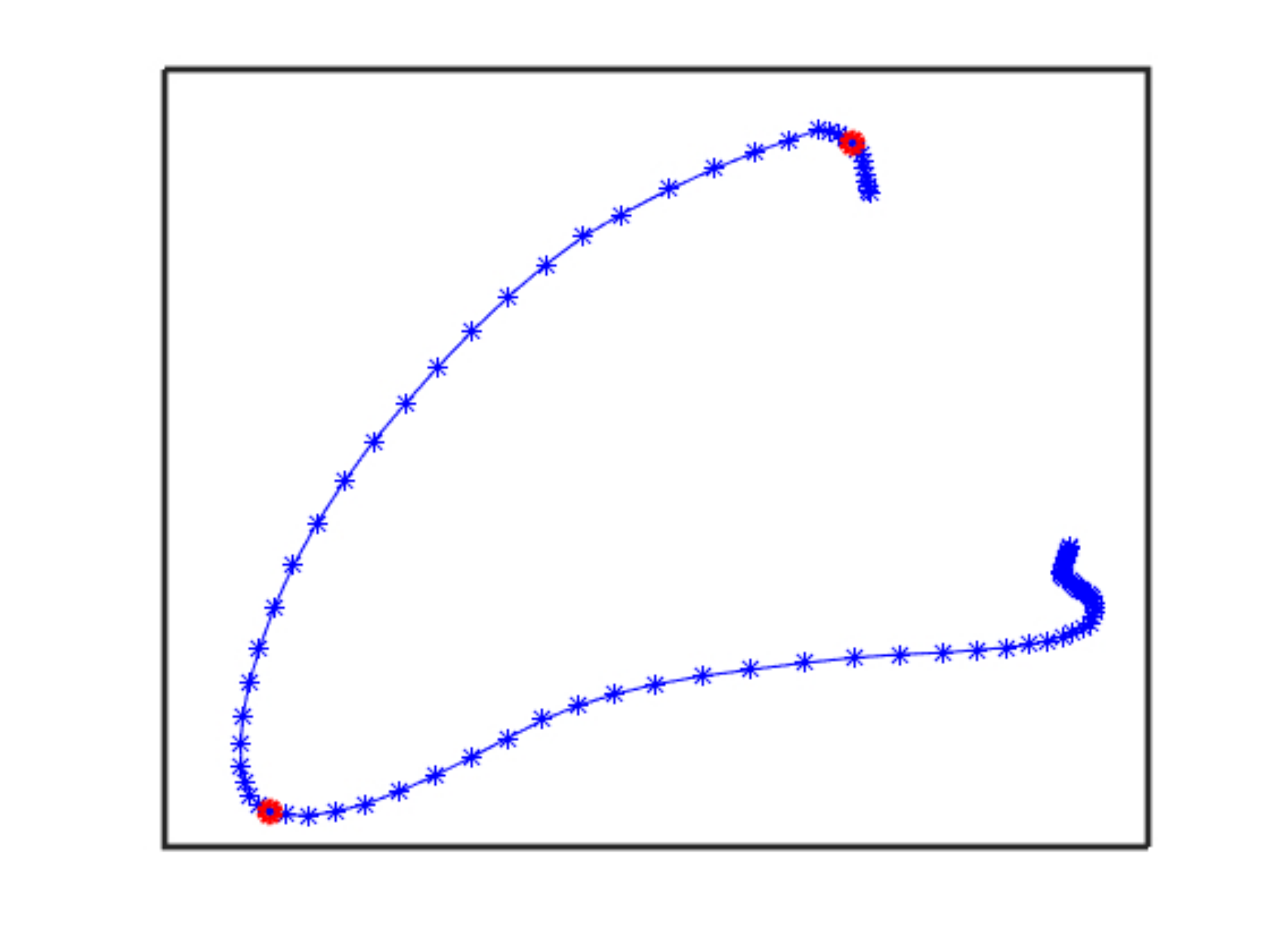}}
\end{tabular}\vspace{-2mm} 
\caption{{Illustration of corner detection results.}}
\label{fig:CD2}
\vspace{-2mm}
\end{figure}

\begin{figure}[t]
\centering
\small
\includegraphics[bb=12 9 420 206, width=0.78\columnwidth]{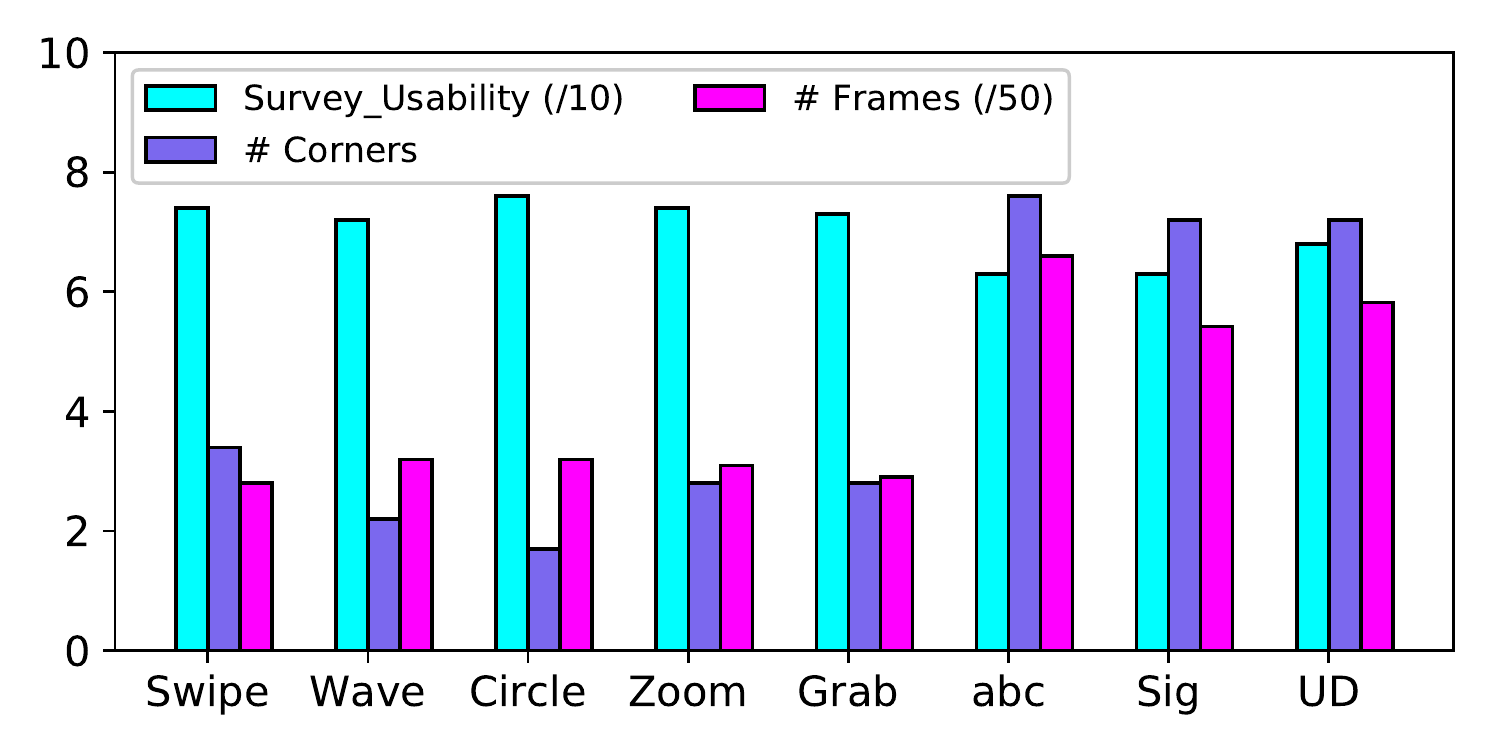} \vspace{-2mm}

\caption{{The usability evaluated based on SUS scores from all participants, the number of corners  and the number of frames from \glab. 
}\vspace{-4mm}}
\label{fig:corner_length_score}
\end{figure}  

\subsubsection{Results from Gesture Samples} To examine the usability of gestures from an objective perspective, we inspect the gesture samples, and calculate the number of corners and the number of frames.

\textbf{Number of Corners.}
We consider the gestures with a larger number of sharp changes more complex, e.g., the gesture $'w'$ is more complex than the gesture $'o'$.  Thus, we use the number of corners of gesture samples to evaluate their complexity. We define the corner of a gesture as the turning point with large curvature value. 
We use a corner detection algorithm based on the curvature scale space (CSS) detector~\cite{CornerHeY04}. We evaluate the gesture trajectory from the index finger, because trajectories from other fingers are similar. 
\fig \ref{fig:CD2} illustrated an example results of the corner detection. 
The slate bars in \fig \ref{fig:corner_length_score} illustrate the average number of corners detected by each gesture type. The gestures \swipe, \wave, \cir, \zoom and \cf are simple ($\#{corners} \leq 3.4$), while the gestures \abc, \ud and \sig are complex ($\#corners \geq 7.2$). 

\textbf{Number of Frames.}
We consider the duration of performing the gesture as a factor to determine whether a gesture is easy to perform or not. A shorter performing duration indicates easier to perform. We define the enrolling time as the number of gesture frames, because the devices' sample rate is stable. The average numbers of frames show similar trends as the number of corners, as shown in \fig \ref{fig:corner_length_score} with pink bars. The simple gestures (\swipe, \wave, \cir, \zoom and \cf) have $\# {frames} \leq 160 $, and the rest of gestures have $\# {frames} \geq 271 $.

From \fig \ref{fig:corner_length_score}, we can see the number of corners and the number of frames exhibit consistent trend. What's more, they are in inverse relationship with SUS scores.
The only exception is \ud . \ud has similar number of corners and frames as the ones of \sig, yet \ud has a higher usability. We believe it is caused by user preferences: users have the full control in choosing a relatively complicate gesture yet they feel easy to perform, unlike all other gestures that are forced upon them.


\vspace{-3mm}
\subsection{Security vs. Usability }
\label{sec:tradeoff}
In this section, we explore the relationship between security and usability based on the survey responses and  quantitative metrics.

\begin{figure}[t]
\centering
\small
\vspace{-0mm}
\begin{tabular}{cc}
\includegraphics[bb=16 4 385 299, width=0.44\columnwidth]{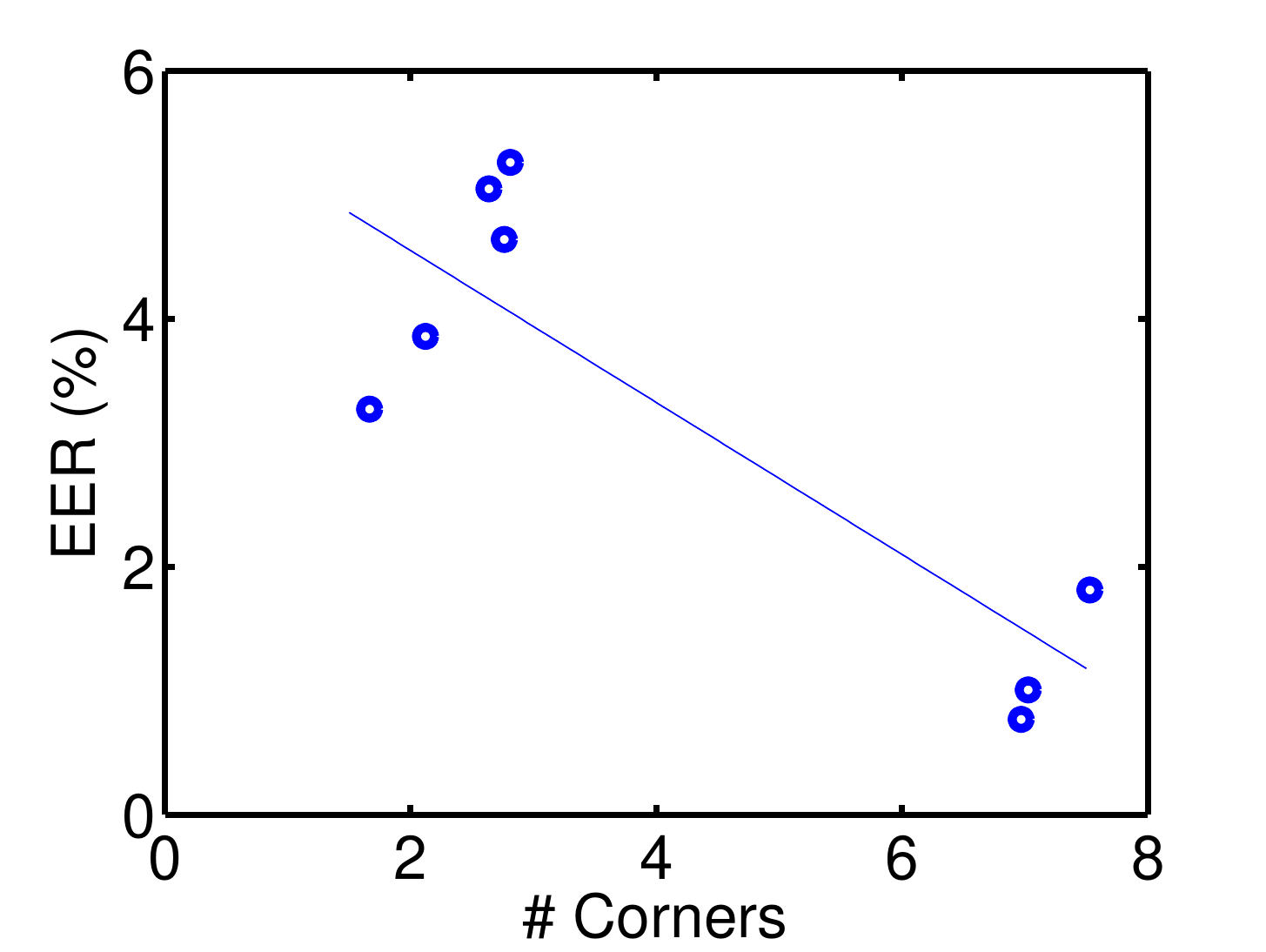} 
&
\includegraphics[bb=16 0 385 299, width=0.44\columnwidth]{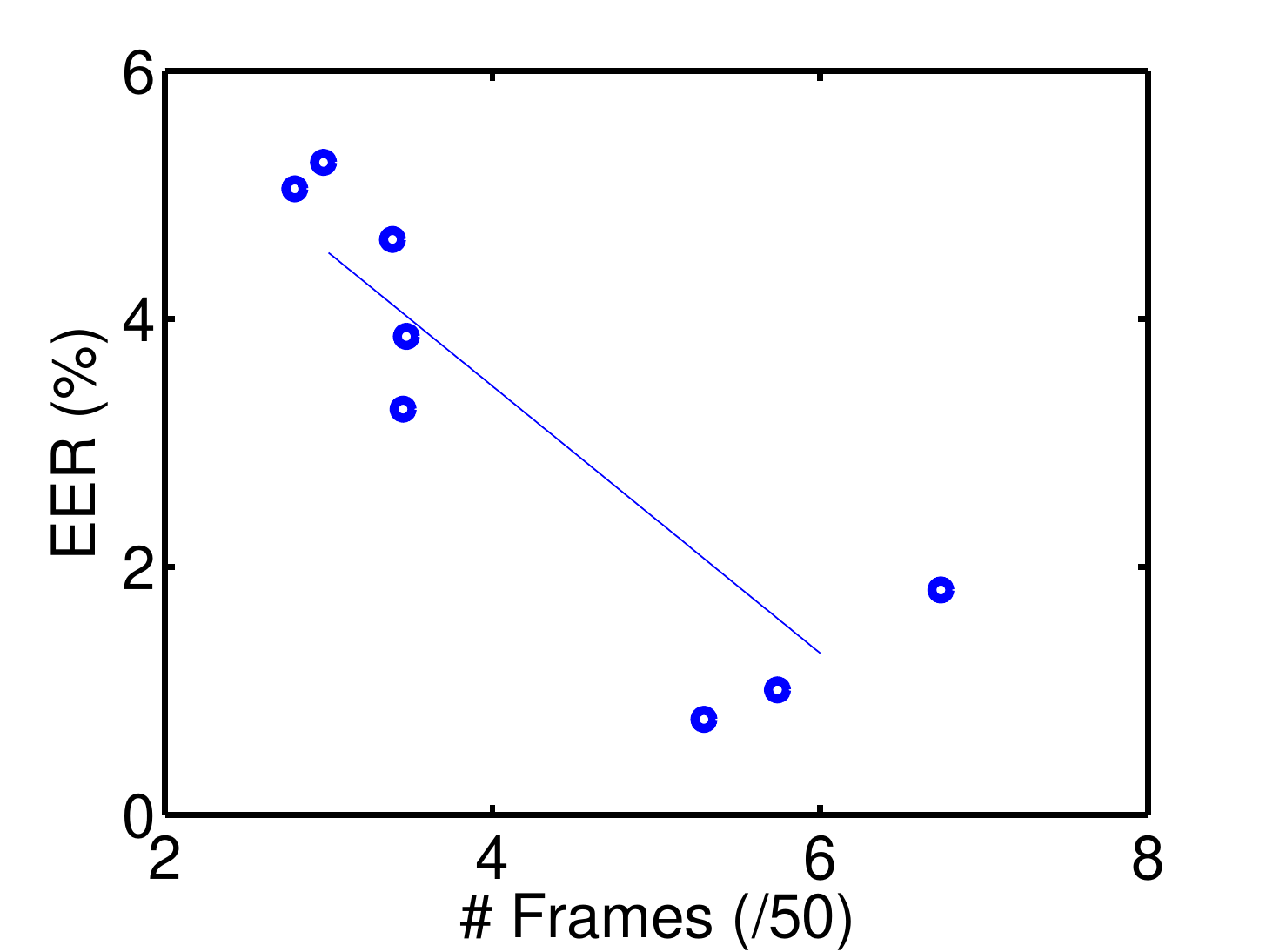} \\
\end{tabular}\vspace{-2mm}
\caption{{The correlation between EER and the number of corners [left] and the number of frames [right]. 
}\vspace{-4mm}}
\label{fig:eercorner}
\end{figure}

\subsubsection{Results from Survey Responses}
To compare the subjective evaluation of usability and security, we use the SUS scores from survey questionnaires as usability metric and use the security scores that are represented by the percentage of participants who consider gestures as ``Most secure'' and ``Second secure'' out of five options (i.e., from least secure to most secure). By scaling all results to the range of 0 to 100, we show the average scores of each gesture from all participants in \fig \ref{fig:usa_sec_survey}. Higher bars indicate better performance for both usability and security. We observed that the participants consider the gestures that are easier to perform (e.g., \swipe, \wave, \cf, \cir, and \zoom) as being less secure, while the gestures \sig, \ud and \abc are considered more secure but scarified some usability.

\subsubsection{Results from Quantitative Metrics}

We used the average number of corners and number of frames as evaluation metrics for usability and EER as a metric for security. To analyze the trade-off between security and usability, we chose the linear least square fitting technique to model the relationship between them.  As a result, we obtained fitting lines with coefficient values $r = -0.85$ (number of corners v.s. EER) and $r = -0.88$ (number of frames v.s. EER), shown in \fig \ref{fig:eercorner}. All gestures roughly follow the inverse relationship between usability and security. That is, with the increase of the number of corners and the increase of gesture length, the security performance improves.  

In addition, the two plots showed the gesture set can be clustered into two subsets. One subset  consists of gestures \abc, \ud, and \sig (i.e., the 3 dots at bottom-right area in both subplots of \fig \ref{fig:eercorner}), which have a larger number of corners and number of frames but lower EER values.  The other subset consists of gestures \swipe, \wave, \cir, \zoom, and \cf,   which have fewer corners and frames, but higher EER values. The existence of two subsets indicates the securer gestures generally do not have better usability in terms of number of corners and gesture length.

\section{Consistency Study}
\label{sec:consistency}

To evaluate the consistency of each gesture, we collected samples over 6 weeks. Participants came and contributed data three times per week for the first two weeks, and twice per week for the next two weeks, and once per week for the remaining weeks. Most participants finished the entire experiment. Only two users quit before the end of 6 weeks due to personal reasons. Table~\ref{tab:data} summarizes the information of each round of data collection. 

\vspace{-3mm}
\subsection{Consistency from Survey Responses}
We asked each participant how likely they could remember the gesture before each batch of gesture collection, except the first collection. Note that all participants can remember how they performed each gesture without any hint at each batch of gesture collection. For each gesture, the memorability responses from the survey also indicate that more than 80\% of participants agree that they can recall it. There was no significant difference among all the gestures.

\begin{figure}[t]
\centering
\begin{tabular}{c}
{\includegraphics[bb=52 3 753 279, width=0.72\columnwidth]{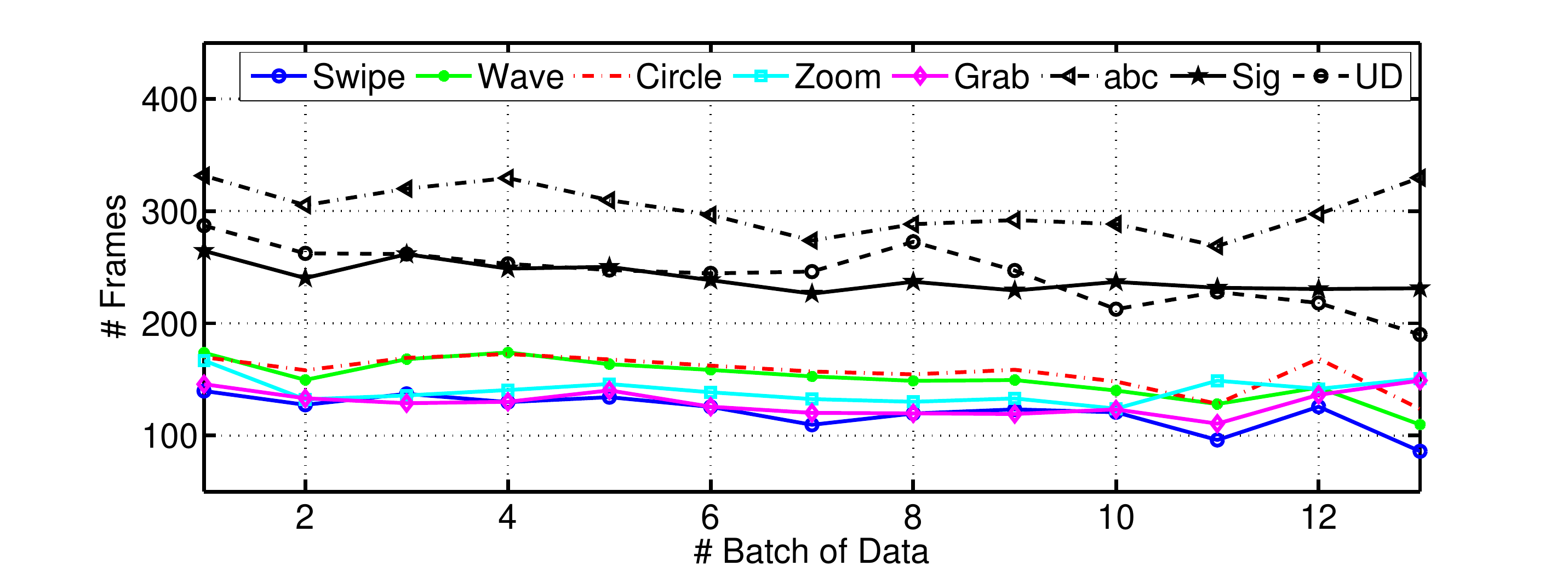}} \vspace{2mm} \\

{\includegraphics[bb=72 3 753 279, width=0.70\columnwidth]{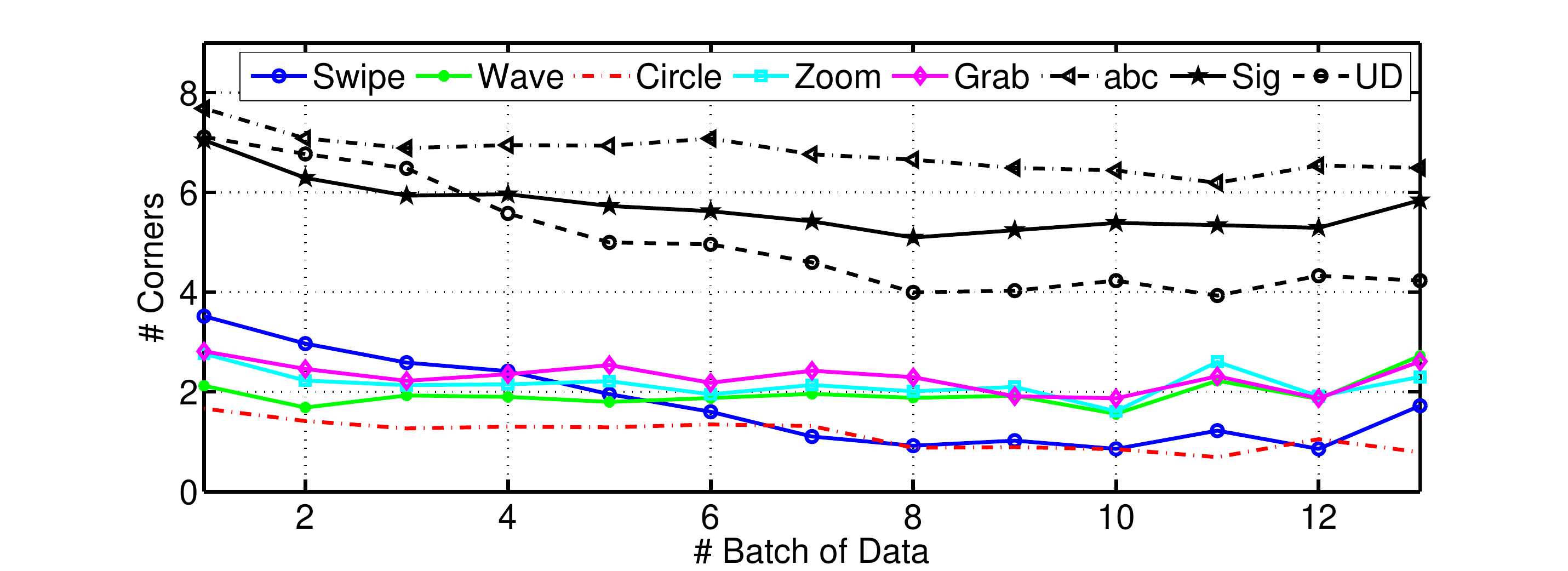}} \\
\vspace{-4mm}
\end{tabular}
\caption{{[Top] The average number of frames and [Bottom] the average number of corners of each batch of data.}\vspace{-4mm}}
\label{fig:enroll_10}
\end{figure}

\vspace{-3mm}
\subsection{Consistency on Gesture Samples}
We examine the consistency of all types of gestures using the two metrics for quantify usability: the number of frames and corners. \fig \ref{fig:enroll_10} shows the average number of frames  and corners by gesture types over 10 batches of collection. 
Overtime, we observed that participants become increasingly proficient with each gestures and thus tend to perform gestures faster and smoother.



\vspace{-3mm}
\subsection{Consistency on EER}
In this section, we use EER to quantify consistency. We consider the EER values obtained when the training samples and testing samples belong to the same batch of data as the baseline. We quantify the changes of two batches of gesture samples by the difference of two EER (i.e., increase of EER) values: between the baseline EER and the EER obtained when using the training samples of one batch and testing samples from the other. A smaller difference indicates a better consistency.  In particular, we ask two questions:  (a) Will gestures performed over time change? (b) Will the change of gestures converge over time? To answer these question, we designed two experiments.

\begin{figure}[!tp]
\centering
\scriptsize
\vspace{-0mm}
\begin{tabular}{c}
\includegraphics[bb=44 9 624 284, width=0.86\columnwidth]{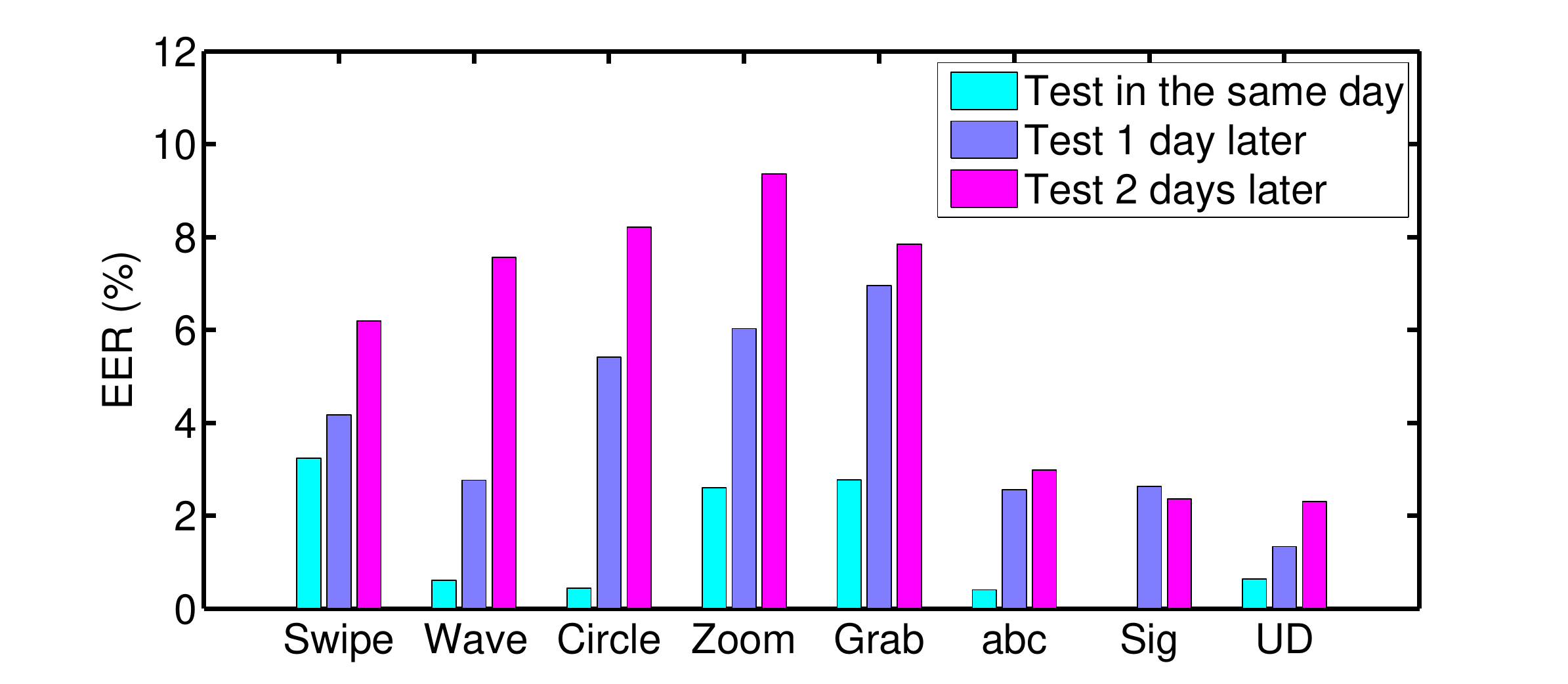} \\
\end{tabular}
\vspace{-1mm}
\caption{{The results of security performance tested by 25 participants one day later and 23 participants two days later.}}\vspace{-2mm}
\label{fig:eer_by_day}
\end{figure}

\textbf{Consistency over Time.} To understand whether the change of gestures is proportional to the gap between performing gestures, we select batches of samples that were collected in two consecutive days and in every other days. As we could not force participants to perform gestures in a tightly-controlled time schedule, we only managed to find $25$ participants who had contributed data in two consecutive days and 23 participants who had contributed data in every other day. From the results shown in \fig \ref{fig:eer_by_day}, we observe that 1) for almost all gestures, the EER increases as the days go by; 2) the gestures \ud, \sig and \abc have relatively smaller increases in EER than the other simple gestures.



\textbf{Convergence of Gesture Changes.}
To understand whether the changes across multiple batches will be reduced over time, we trained SVM classifiers with the samples from the $1$st batch and tested with the ones from the 1st (excluding the training samples), 2nd, 3rd, 4th, and 5th, etc. 

\begin{figure}[!tp]
\centering
\vspace{-0mm}
\begin{tabular}{c}
{\includegraphics[bb=62 3 753 284, width=0.86\columnwidth]{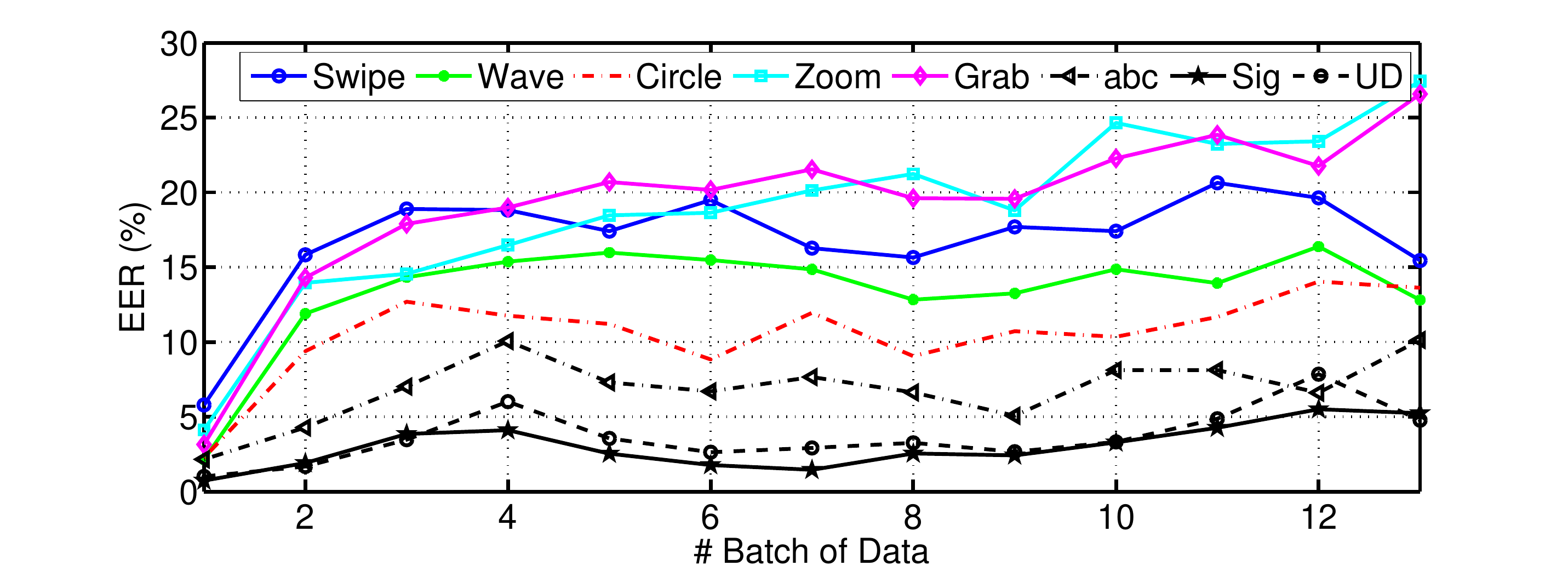}} \\
\end{tabular}\vspace{-0mm}
\caption{{The EER results averaged over each type of gesture.  The experiments are based on the training templates from the first batch of data. }}\vspace{-4mm}
\label{fig:eer_train1}
\end{figure}

The results are shown in \figref{fig:eer_train1}. We have the following observations. 
\begin{itemize}
 \item The EER results are low when tested in the same day, and increase fast at the first gap (two days on average), then show convergence around the 5th batch (10 days on average). 
\item \sig and \ud exhibit the best security performance ($EER < 6\%$ for all batches). The reason could be that they are complex gestures, and the relative changes in terms of the number of corners and frames are smaller than the other gestures. 
\item  \abc presents relatively better performance than the rest of simple pre-defined gestures. It is a complex gesture like \sig but every participant performed the same gesture, leaving little space to tolerant changes. 
\item  \cir has medium performance. \cir does show less changes of the number of corners than gesture \swipe or \zoom. 
\item \swipe, \cf, \zoom, and \wave have the worst performance. 

\end{itemize}

\begin{figure}[t]
\centering
\begin{tabular}{c}
{\includegraphics[bb=31 4 537 254, width=0.85\columnwidth]{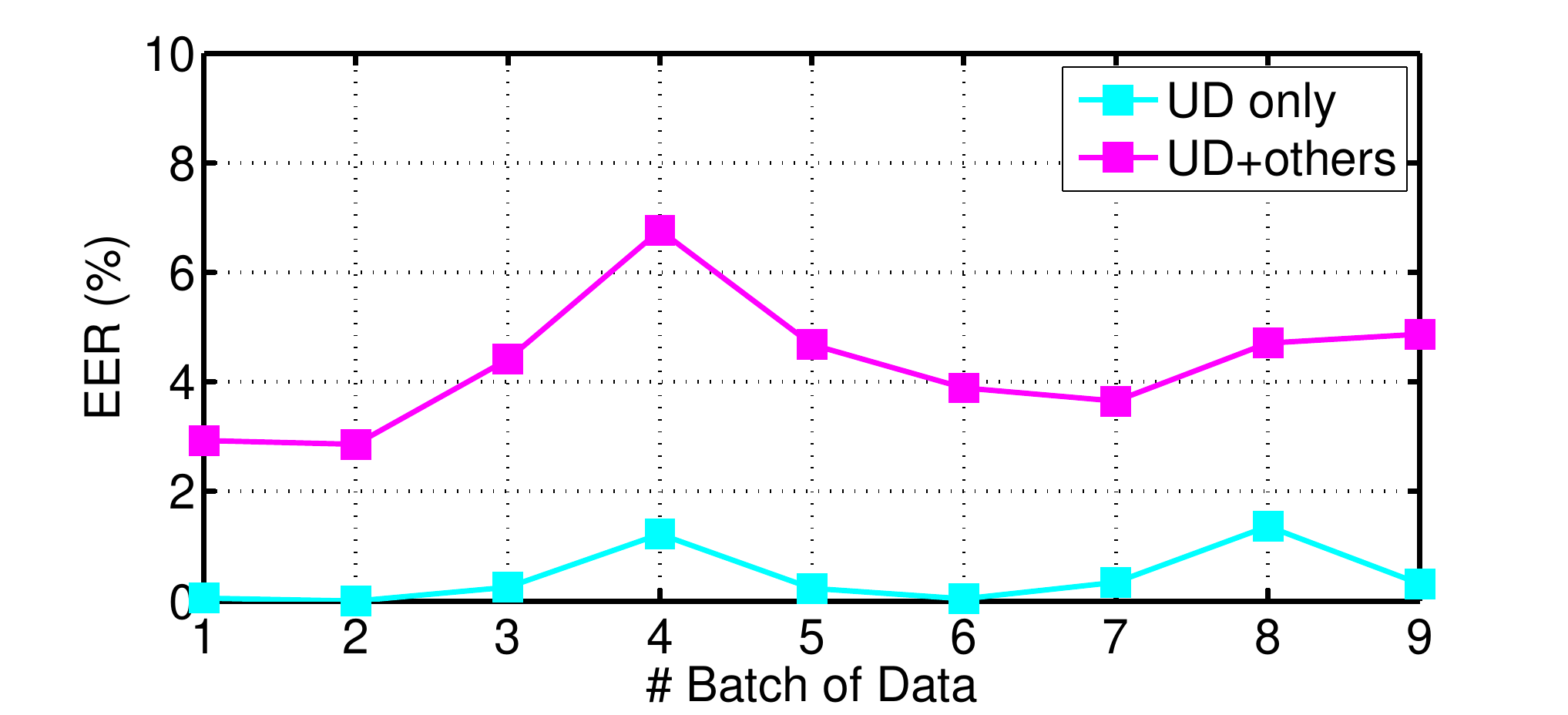}} \\
\end{tabular}\vspace{-1mm}
\caption{{The average EER results of \ud gesture using only the first batch of data for training. `UD only' shows the average results of 10 participants. `UD+others' shows the average results of the other 32 participants.}}\vspace{-4mm}
\label{fig:ud_only}
\end{figure}

\textbf{The Number of Gestures to be Remembered.} Our experiments involved two groups of users: 32 participants that required to remember and perform all gestures and 10 participants that only needed to perform \ud gestures. The latter group mimics reality where users choose a few gestures as passwords. To understand the difference between two groups, we trained SVM classifiers with the \ud samples from the $1$st batch and tested with the ones from the 1st (excluding the training samples), 2nd to 9th batch of data. The results are shown in \fig \ref{fig:ud_only}, from which we observed that  without the burden of remembering other gestures, the 10 participants can remember the gestures better and their gestures over time exhibited much better distinctness and consistency.



\section{Shoulder Surfing Attacks}
\label{shouldersurfing}
It is unclear whether mid-air gestures are resilient to shoulder surfing attacks. To gain insight of shoulder surfing, 
We recruited 13 subjects as victims and another 4 subjects as attackers, who mimic each type of the gesture performed by victims. Each victim enrolled  12 samples for each type of the gestures studied in this paper, and thus $13 \times 132 \times 8$ victim samples are collected.  
To mimic shoulder surfing attacks, we record short videos (e.g., one or two gesture instances) while victims are performing the gestures. 
\begin{figure*}[!tbph]
\scriptsize
\centering
\vspace{-0mm}
\begin{tabular}{cccc}
\subfigure[\swipe]
{\includegraphics[width=.47\columnwidth]{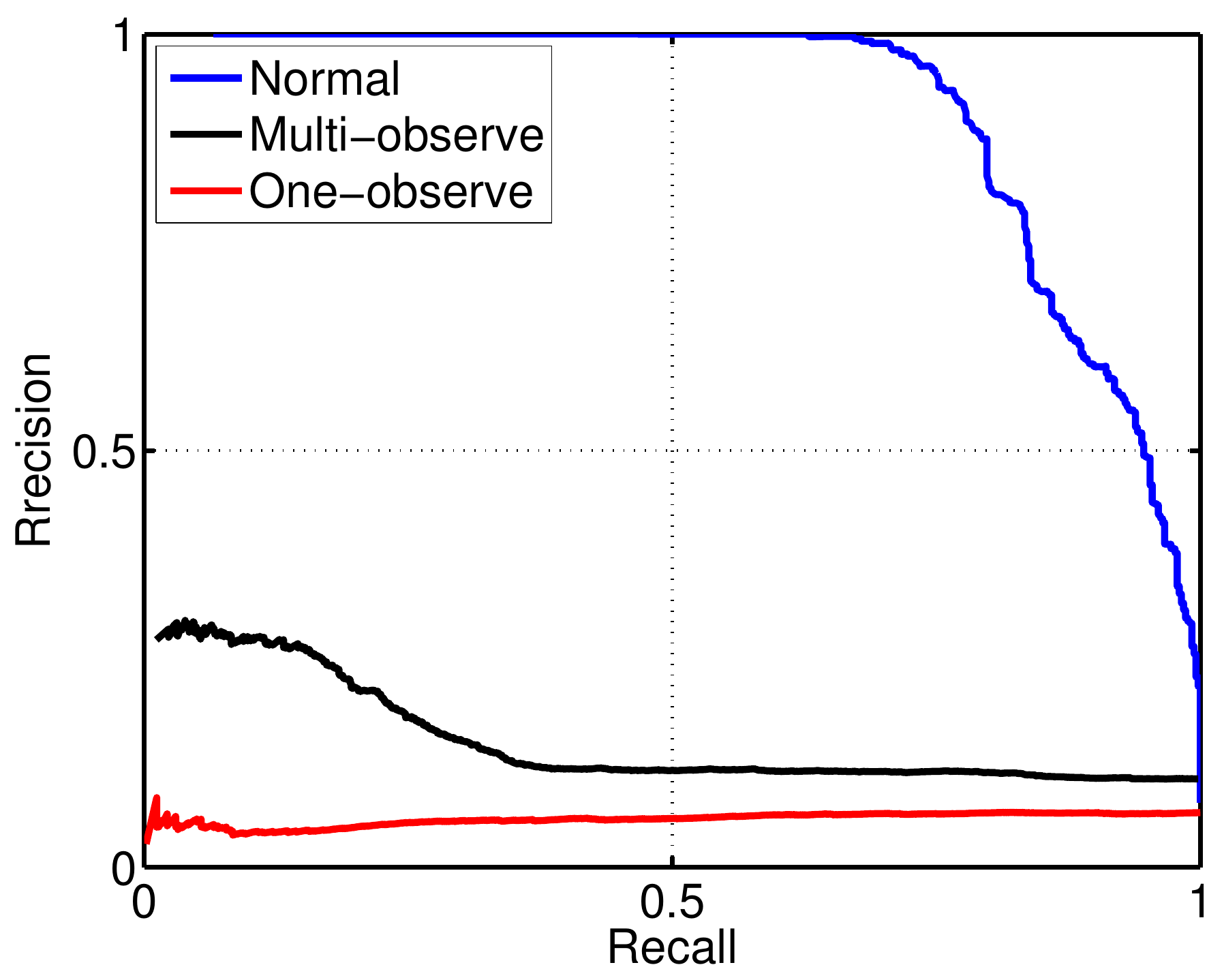}}
&\subfigure[\wave]
{\includegraphics[width=.47\columnwidth]{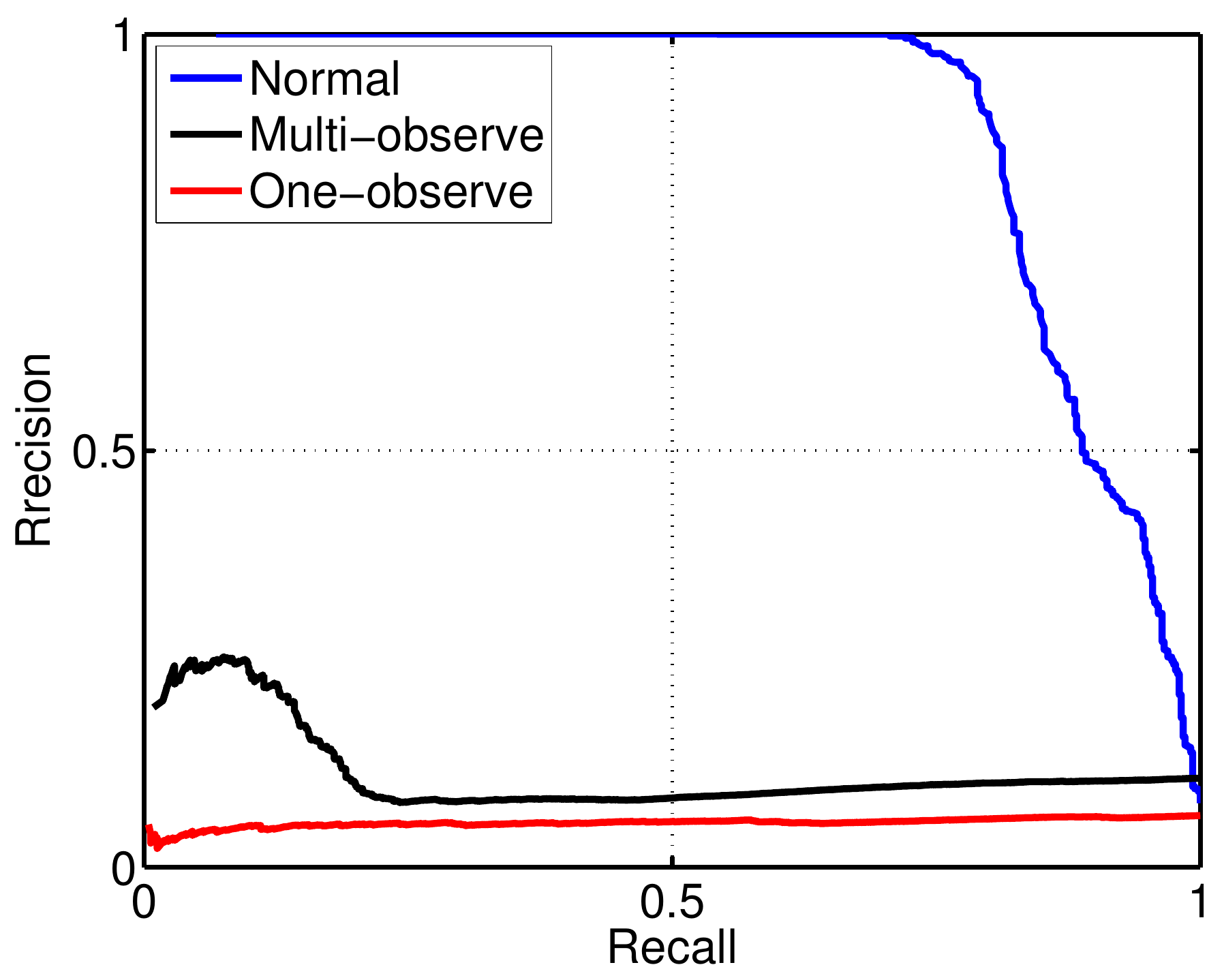}}
&\subfigure[\cir]
{\includegraphics[width=.47\columnwidth]{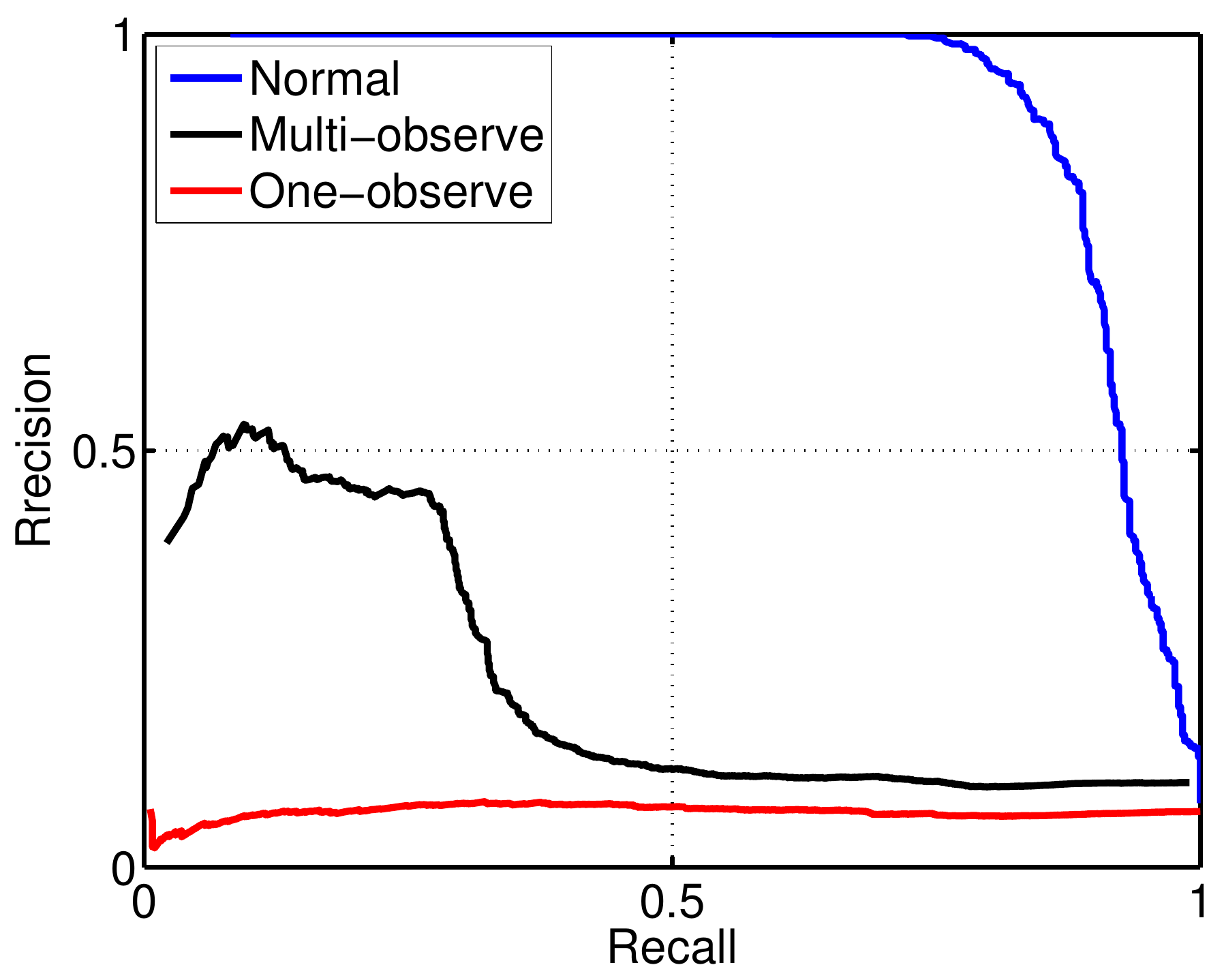}}
& \subfigure[\zoom]
{\includegraphics[width=.47\columnwidth]{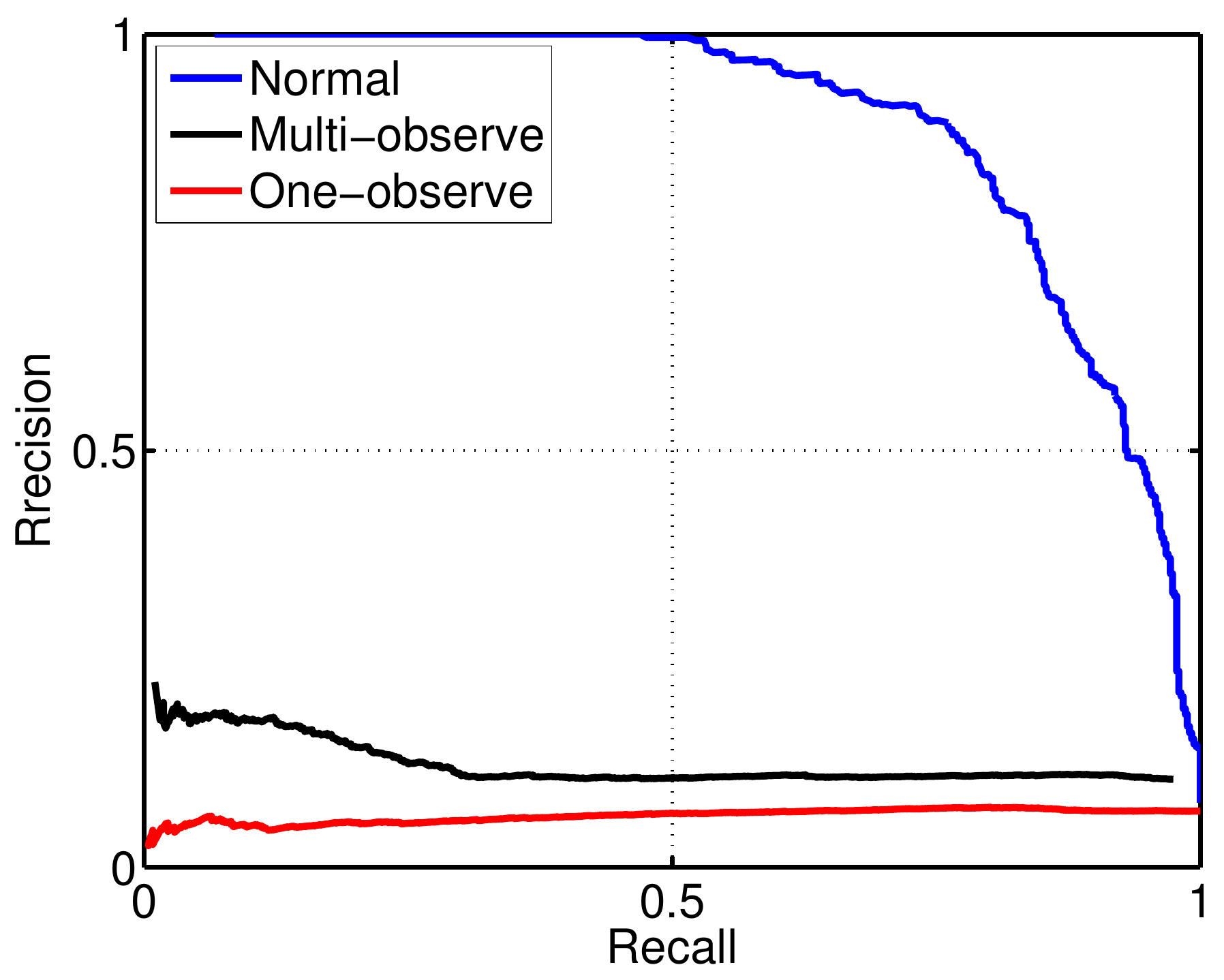}} 
\\
\subfigure[\cf]
{\includegraphics[width=.47\columnwidth]{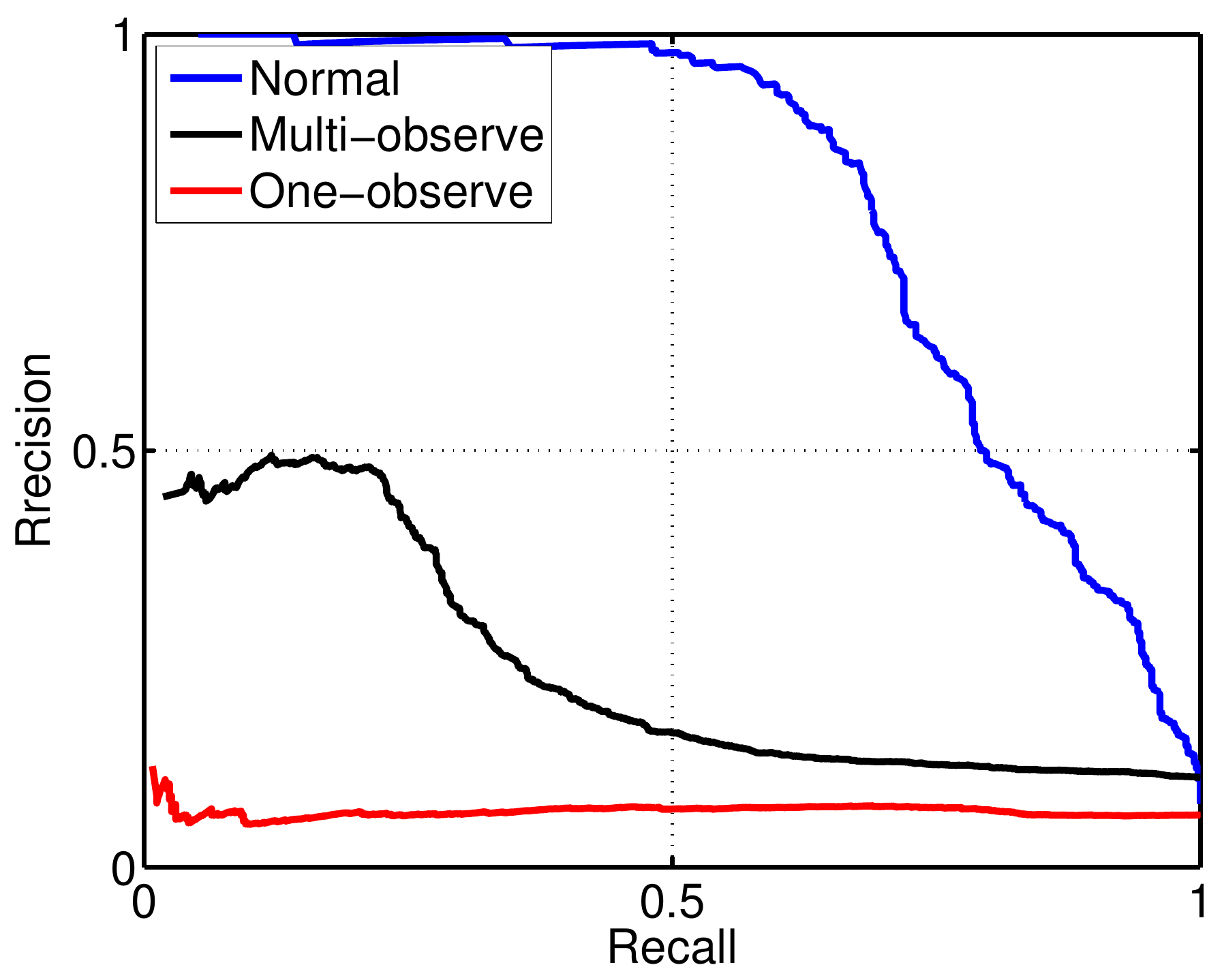}} 
& \subfigure[\abc]
{\includegraphics[width=.47\columnwidth]{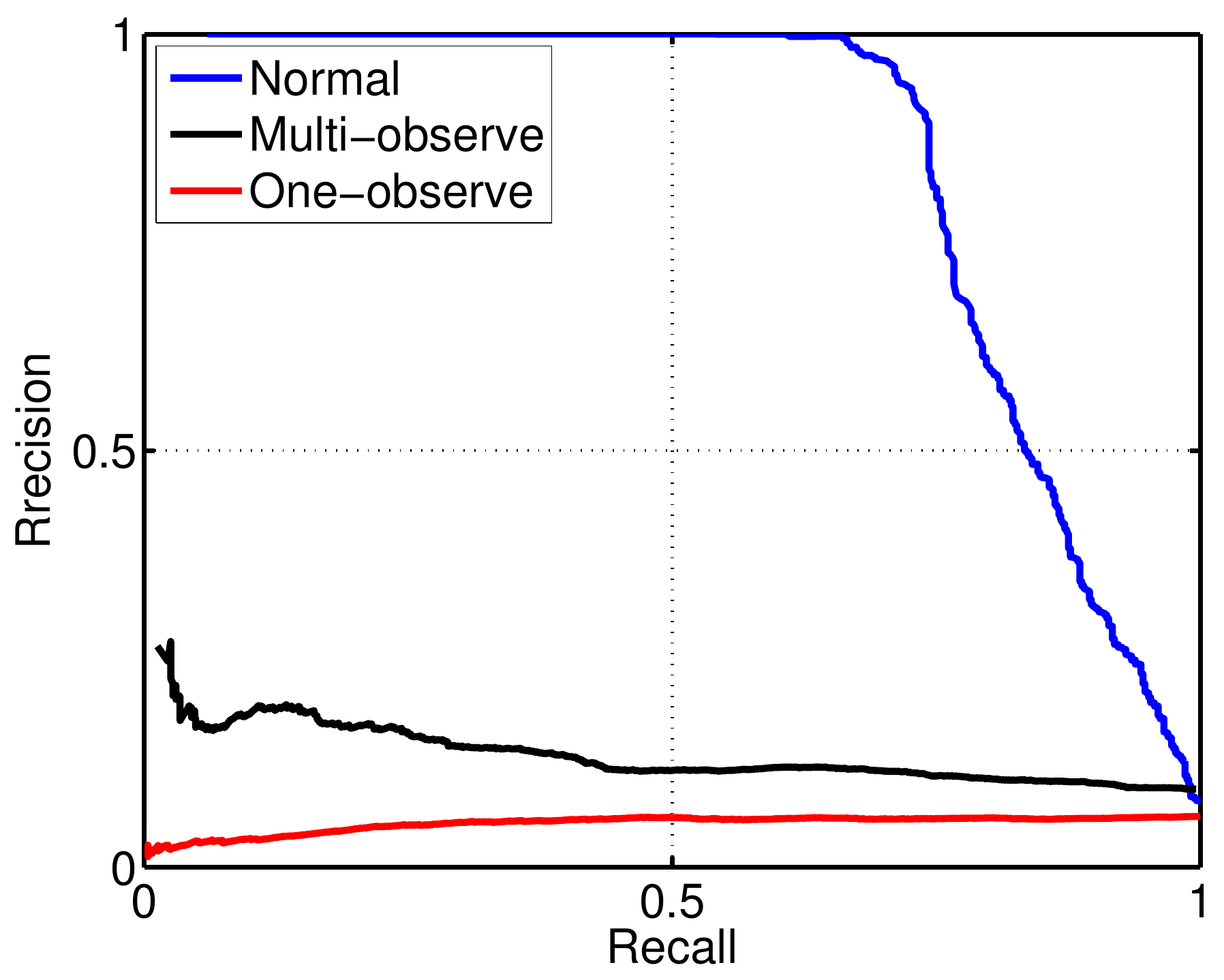}} 
& \subfigure[\texttt{sig}]
{\includegraphics[width=.47\columnwidth]{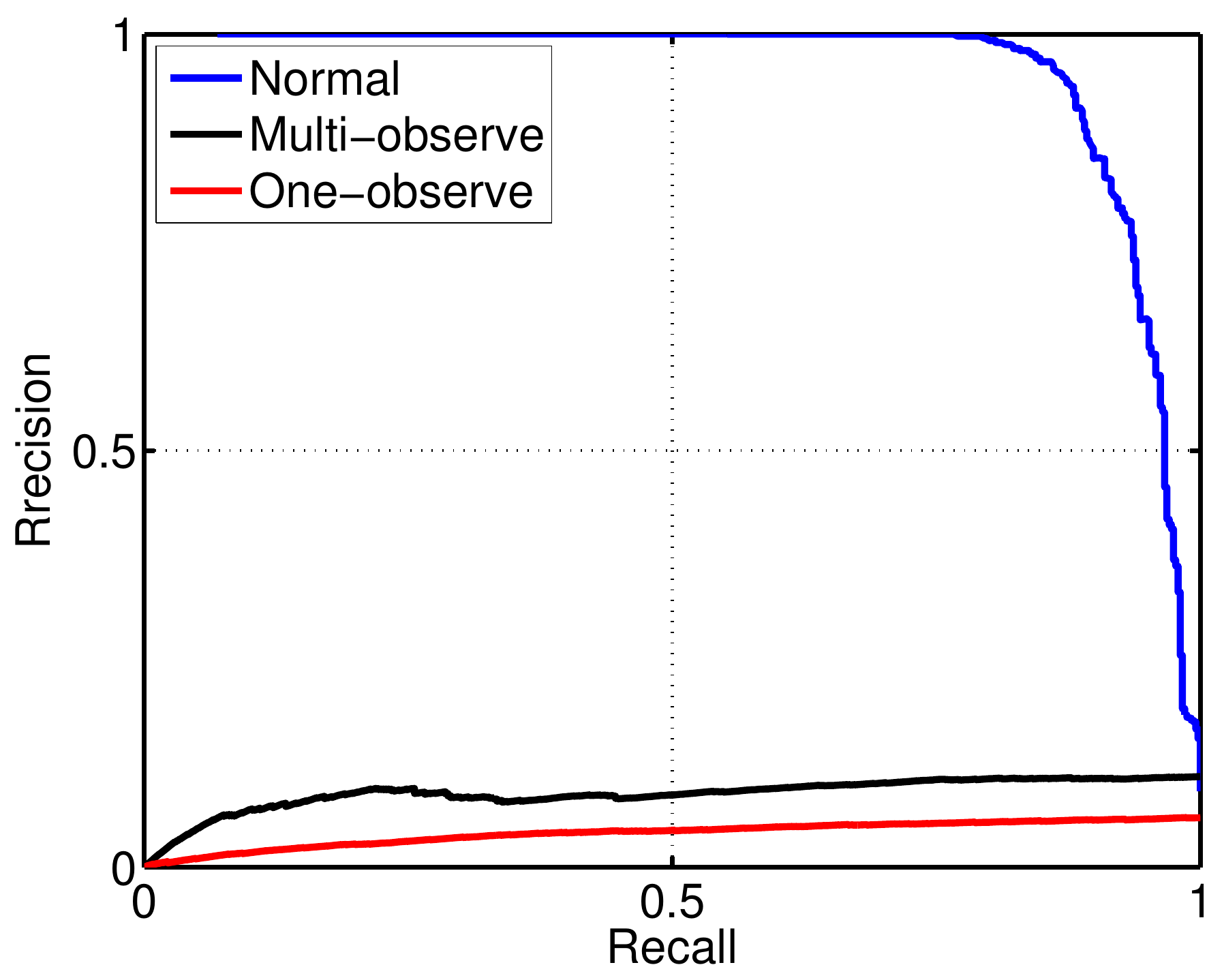}}
& \subfigure[\ud]
{\includegraphics[width=.47\columnwidth]{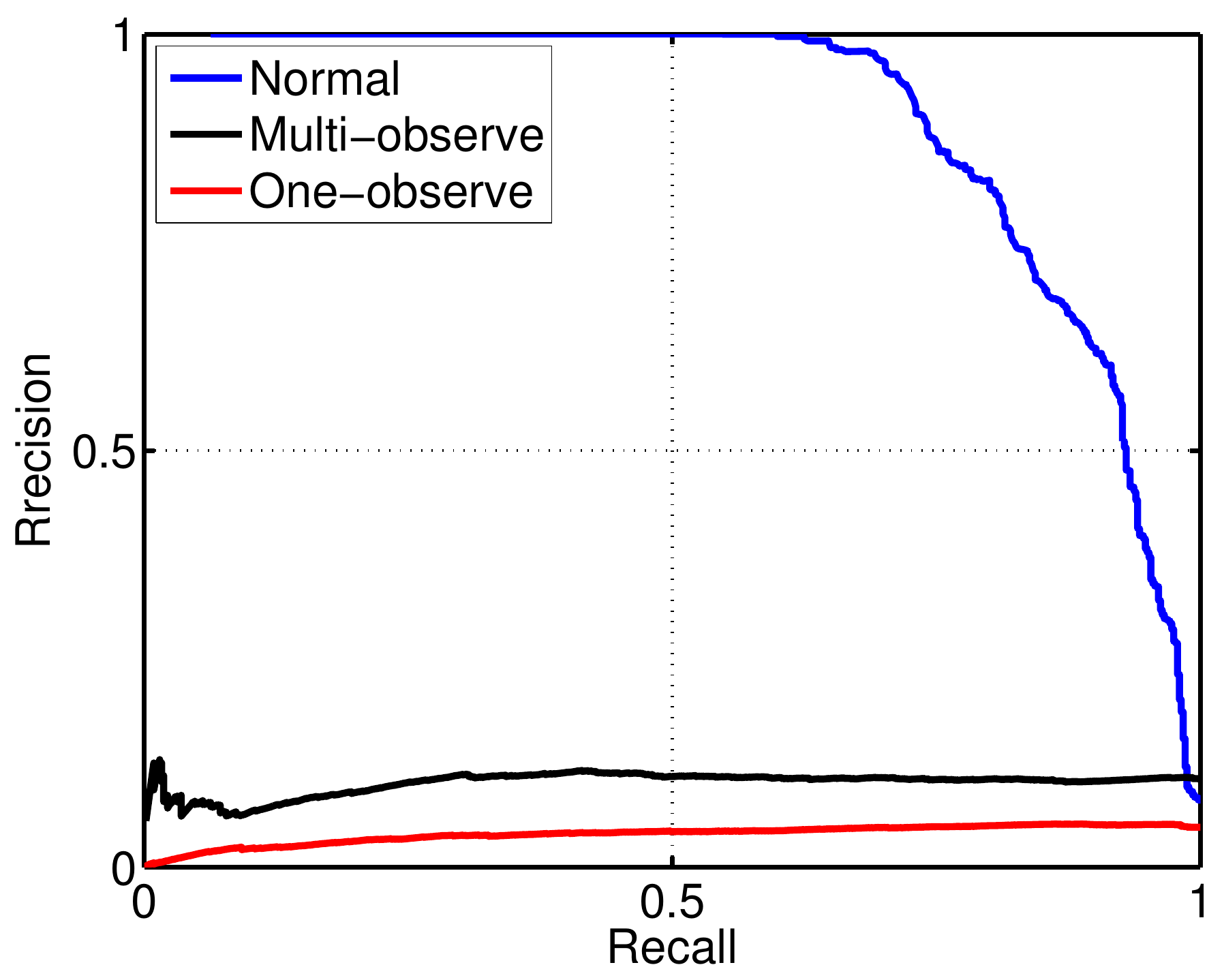}}
\end{tabular}\vspace{-2mm} 
\caption{{precision recall curves under normal (blue) and attack scenarios. In general, the attack with multiple times observation has better performance than one time observation attacks.}}
\label{fig:sf}
\vspace{-4mm}
\end{figure*}

With the videos, we consider two attack scenarios. In the first scenario, the attackers are allowed to watch the videos only once, representing the case that  they may happen to see the victim entering gestures once. 
Then, the attackers entered five gestures by mimicking what they saw. Specifically, they try 5 times for each shown gesture. In total, $4 \times 5 \times 8$ attack samples are collected. 
In the second scenario, the attackers can watch the videos as many times as they want before or during the attack. Each attacker attacks $X=$ 10 to 15 times while learning from a recorded video. In total, we have $4 \times X \times 8$ attack samples. 

To evaluate both attack scenarios, we tested the attack samples with the classifiers trained with victim samples (used 4 samples for each class). For comparison, we also tested the classifiers with victim samples which are not used for training. 
We use precision recall curves to illustrate the results. 
By varying the threshold to reject possible impostors, we obtain precision recall curves that indicate the trade-off between security and usability.  A higher precision indicates a more strict threshold (i.e., better security), at the cost of letting legitimate users try more times. A higher recall indicates a less strict threshold and may let some attackers pass authentication, but legitimate users could pass authentication with a less number of attemps. 

The upper-right corner of a curve is the idea point (i.e., 100\% precision and 100\% recall -- all legitimate users are authenticated with one attempt, and all the attackers are rejected).  
\fig \ref{fig:sf}  shows precision recall curves of each gesture types with different types of test sets: normal samples from victims (depicted in blue), attack samples with multiple times of learning (black), attack samples with once observing (red).
 Results of all types of gestures show that attack samples with one or multiple time observation both have low precision, although multiple learning did slightly improve the chances of attacks. Nevertheless, the precision and recall are still relatively low. Thus, the mid-air gestures are difficult to mimic and the shoulder surfing is not a main thread to our authentication system.  

\section{Related Work}
\label{related}

Gestures, as a new way of human computer interaction, have shown great promises, and an extensive literature on gesture recognition exists, which includes multi-touch pinch gestures~\cite{Hoggan:2013}, 3D gesture recognition using accelerometer and gyro~\cite{BLSTM-RNN:2015}, multi-layer gesture recognition with Kinect~\cite{Jiang2014}, and air gesture identification~\cite{Segmentation:2014}. These gestures have been applied to a wide variety of fields, ranging from controling robots~\cite{IJARAI:2013}, computer commands~\cite{chen2014airlink}, authentication purposes~\cite{ahmed2015checksum, Uell:CCS13, shresthacurbing}, game control~\cite{kinectgame} to VR commands~\cite{Kulshreshth:2014}.



An increasing number of studies focus on user authenticating based on behavioral biometrics. Such gestures are embedded in the usage pattern of traditional I/O devices, such as keystroke dynamics and mouse movement patterns~\cite{Monrose:CCS99,Jorgensen11mouse}.
With the emerge of new technology (e.g., sensors or touch screens), new gestures were discovered. Lower leg gaits~\cite{accelerameter_shaking_2006} and hand gesture patterns~\cite{accelerameter_shaking_2006}
captured by accelerometers have shown to achieve high accuracy in user authentication.  
The operations on a smartphone/pad's touch screen (e.g., writing a word or using an unlock pattern) can be used to authenticate users either once during logging in~\cite{Uell:CCS13} or continuously thoughout the oepration~\cite{LiZX:NDSS13}. The security and memorability of multi-touch gestures for mobile authentication have been studied~\cite{sherman2014user}.
Unfortunately, touch gestures can be vulnerable to shoulder surfing or smudge attacks~\cite{Aviv:woot10}. 

Mid-air gestures have become a hot topic recently. 3D hand gesture has been studied on touch-less interactions, such as augmented reality application and game-based virtual environments~\cite{TarantaII:2015:EBC, 3d_gesture_control, vision_ges_recognition, VR_touchless_interaction, virtual_manipulation}. For authentication purpose, Nigam~\etal combined signature gesture captured by Leap Motion and facial information by a RGB camera to authenticate a user~\cite{Nigam15:LeapSigVeri}. Aslan \textit{et al.} explored two mid-air gestures for authentication in different situations~\cite{Midair_Authentication_Gestures_ICMP14}. AirAuth system evaluated the security performance with a set of simple hand gestures captured by a depth sensor Creative Senz3D~\cite{airauth_chi14_abstract}.
Using Micrsoft Kinect, Hayashi~\etal utilized fusion data of hand waving gesture and user's body length for authentication~\cite{CHI2014_wavetome}, and Tian~\etal used a gesture of whiting signatures in the air~\cite{TianQXW13:NDSS13} for user identification. This paper also investigate mid-air hand gestures, but focuses on quantifying usability and security of various gestures and exploring their relationship. 

Usability evaluation of authentication schemes for other purposes, e.g. password usage in daily life, Touch-ID on iphone, Biometric authentication on smartphones, is a well-researched area~\cite{BioAuthentic, TouchID, diaryStudy}. Usability is crucial for an authentication system to be adopted by users~\cite{undercover}. Although there are many papers on mid-air gesture-based authentication, they mostly focus on improving the accuracy. Only a few literature explored one or two aspects of usability surrounding gesture-based authentication: BroAuth~\cite{Maurer:2012} present an authentication mechanism based on body gestures. They evaluate the usability and security of three types of visual feedback and found that an abstract representation is the best trade-off between security and usability. Aslan \textit{et al.}~\cite{Midair_Authentication_Gestures_ICMP14} studied 13 participants' perceptions on two authentication gestures from the prospective of their emotions. AirAuth~\cite{airauth_chi14_abstract} compared participants' pleasantness and excitement level between a set of predefined gestures. They found a positive correlation between the authentication accuracy and participants' excitement and pleasantness. This paper is along the same line, but aims at discovering general metrics that can quantify both usability and security, and to understand the relationship between usability and security for guiding gesture evaluation. 

\section{Conclusion}
\label{sec:conclusion}
This paper studied the usability and security of a collection of mid-air gestures as biometrics for authentication. 
Through a user study that engaged 42 participants to collect gesture samples 13 times over a 6-week period and a survey, we managed to validate that the quantitative metrics (i.e., the number of corners, the sample frame length, and the equal error rate (EER)) confirms with the subjective scores from the user survey. Further, we find the correlation between security and usability metrics, which shows that an easy-to-use gesture generally has a worse security. Thus, we can utilize the number of corners and the sample frame length to quickly quantify the security of a gesture. Finally, our consistency study shows that participants tend to forget gestures between experiment rounds.


We note that our experiment results may be different from reality because samples were collected in a lab environment, where no serious consequence will be incurred if a user cannot pass the authentication in our study. We envision that after gesture-based authentication is widely used, the inconsistency of gestures over time  will become smaller because we have observed in our study that repeatedly performing gestures will help to provide consistent gestures. 
\bibliographystyle{IEEEtran}
\bibliography{Sections/ref}


\end{document}